\documentclass[useAMS,usenatbib]{mn2e}
\usepackage{amssymb, latexsym, amsmath, graphicx, url, lscape}
\usepackage{xcolor}

\newcommand{\ao}{ApOpt}

\newcommand{\apjl}{ApJ}
\newcommand{\apjs}{ApJS}
\newcommand{\aap}{A\&A}

\title{A Massive Cluster at $z=0.288$ Caught in the Process of Formation: The Case of Abell 959}
\author[L.~B\^{\i}rzan et al.]{L.~B\^{\i}rzan$^{1}$\footnote{E-mail address: lbirzan@hs.uni-hamburg.de},
D.~A.~Rafferty$^{1}$,  R.~Cassano$^{2}$, G.~Brunetti$^{2}$, R.~J. van~Weeren$^{3}$,
\newauthor
M. Br\"{u}ggen$^{1}$,
H.~T.~Intema$^{3,4}$, F.~de~Gasperin$^{1}$, F.~Andrade-Santos$^{5}$,
A.~Botteon$^{2,6}$,
\newauthor
 H.~J.~A. R\"{o}ttgering$^{3}$, T.~W.~Shimwell$^{7}$
\\
$^{1}$Hamburger Sternwarte, Universit\"{a}t Hamburg, Gojenbergsweg 112, 21029, Hamburg, Germany\\
$^{2}$INAF-Istituto di Radioastronomia, via P. Gobetti, 101, I-40129, Bologna, Italy\\
$^{3}$Leiden Observatory, Leiden University, Oort Gebouw, P.O. Box 9513, 2300 RA Leiden, The Netherlands\\
$^{4}$International Centre for Radio Astronomy Research -- Curtin University, GPO Box U1987, Perth, WA 6845, Australia\\
$^{5}$Harvard-Smithsonian Center for Astrophysics, 60 Garden Street, Cambridge, MA 02138, USA\\
$^{6}$Dipartimento di Fisica e Astronomia, Università di Bologna, via P. Gobetti 93/2, I-40129 Bologna, Italy\\
$^{7}$ Netherlands Institute for Radio Astronomy (ASTRON), P.O. Box 2, 7990 AA Dwingeloo, The Netherlands}

\begin{document}

\maketitle

\begin{abstract}

The largest galaxy clusters are observed still to be forming through major
cluster-cluster mergers, often showing observational signatures such as radio
relics and giant radio haloes. Using LOFAR Two-meter Sky Survey data, we present
new detections of both a radio halo (with a spectral index of
$\alpha_{143}^{1400}=1.48^{+0.06}_{-0.23}$) and a likely radio relic in Abell
959, a massive cluster at a redshift of $z=0.288$. Using a sample of clusters
with giant radio haloes from the literature (80 in total), we show that the
radio halo in A959 lies reasonably well on the scaling relations between the
thermal and non-thermal power of the system. Additionally, we find evidence  that steep-spectrum haloes tend to reside in clusters with high X-ray
luminosities relative to those expected from cluster $LM$ scaling relations,
indicating that such systems may preferentially lie at an earlier stage of the
merger, consistent with the theory that some steep-spectrum haloes result from
low-turbulence mergers. Lastly, we find that halo systems containing radio
relics tend to lie at lower X-ray luminosities, relative to those expected from
cluster $LM$ scaling relations, for a given halo radio power than those without
relics, suggesting that the presence of relics indicates a later stage of the
merger, in line with simulations.

\end{abstract}


\section{Introduction}\label{S:intro}

In the present-day Universe, many clusters are still forming through
hierarchical processes and major merger events with neighboring clusters
\citep[e.g.,][]{pres74,spri06,krav12}. On smaller scales, non-gravitational
processes, such as radiative cooling, supernova heating, and feedback from
active galactic nuclei (AGN), are also important
\citep{bens03,scan04,birz04,voit05,mcna07,fabi12,alex12}. As such, clusters of
galaxies have wide-ranging astrophysical applications. For example, they can be
used to constrain the cosmological parameters \citep{alle11} and to provide
constraints on the properties of dark matter
\citep{marke04,clow04,clow06,harv15}.

All of the processes important to the formation of clusters dissipate energy
into the intra-cluster medium (ICM) through shocks: e.g., accretion shocks,
merger shocks, AGN related shocks, or ICM bulk motion shocks \citep[see the
reviews of][]{brug12,brun14}. Observationally, merger shocks have been detected
in \emph{Chandra} X-ray and XMM-\emph{Newton} observations of a small number of
merging clusters \citep[e.g., Bullet Cluster, A520, A521, A2146, A3667, A754, El
Gordo, A665, A2219, and
A2744;][]{mark02,shim15,mark05,giac08b,bour13,russ10,fino10,sara16,maca11,
bott16a,dasa16,cann17,ecke16,pear17} with modest Mach numbers of $M=1.5-3$ and,
in radio images, in the form of large-scale, diffuse emission associated with
the shocks \citep[e.g.,][]{shim14,bott16a,vacc14,giac08b,golo18}. Such radio
structures, known as \emph{radio relics} \citep[or radio shocks, see the review of][]{vanW19}, have polarized emission resulting from
ordered magnetic fields aligned by the shock. The favored mechanism for the
relic creation is the acceleration of electrons by diffuse shock acceleration
(DSA), where the electrons can either come from the thermal pool
\citep[e.g.,][]{enss98,pfro08} or be mildly relativistic cosmic rays \citep[CRe;
e.g., fossil electrons from previous AGN or merger
activity,][]{mark05,kang12,pinz13,vanW17b}.

In addition to radio relics, a number of luminous X-ray clusters show diffuse,
cluster-scale radio emission known as \emph{giant radio haloes}
\citep{vent07,vent08,kale13,kale15}. The giant radio haloes (RHs) are thought to
form from the post-merger turbulence of seed suprathermal CRe \citep[e.g.,turbulent re-acceleration
model;][]{brun01,petr01,cass06,cass07,brun09,cass10a,brun11,donn13,brun16,pinz17,ecke17,brun17}.
In support of this scenario, there exists a connection between
the presence of a halo and the presence of merging activity, with the radio
luminosity of the halo ($P_{1.4{\rm GHz}}$) correlating with the X-ray
luminosity of the cluster ($L_{\rm  X}$), the mass of the cluster ($M$) or the
Sunyaev-Zel'dovich signal ($Y_{\rm {SZ}}$), albeit with large scatter
\citep{cass07,brun07,brun09,cass10,basu12,cass13,kale15}.

Giant radio haloes (RHs) are generally observed in clusters whose X-ray gas has
a long central cooling time, $t_{\rm cool} > 10^{9}$ yr. These clusters are
known as non-cooling flow clusters (NCF), whereas those with shorter cooling
times are known as cooling flow (CF) or cool-core (CC) clusters. In NCFs, the radio
power of the central radio source is typically below $L_{\rm{1.4GHz}}$ $<$ 2.5
$\times$ 10$^{30}$ erg s$^{-1}$ Hz$^{-1}$ \citep{birz12}. However, there are a
few systems known to have a short central cooling time and to possess a giant
radio halo \citep[e.g., EL Gordo, H1821+643;][]{birz17,lind14,russ10,bona14a}.
There are also systems which are seen to have an intermediate (or large) cooling
time and a two-component RH: a RH plus a radio mini-halo \citep[e.g.;
RXJ1347.5-1145, A2319, A2142, RXJ1720.1+2638,
PSZ1G139.61+24;][]{ferr11,stor15,vent17,savi18a,savi19}. The details of how CF
and NCF systems form and relate to each other are still not fully understood
\citep[e.g.,][]{pool08,burn08,parr10,pfro12,rasi15,hahn17,mede17}, and the
observational bias of X-ray selected samples complicates the issue
\citep{ross17,andr17}.

Giant RHs and radio relics are found in a significant percentage of massive clusters
\citep[e.g., $\sim$ 23\% for EGRHS,][]{kale13}. Therefore, to date, most radio
campaigns searching for such RHs have focused on luminous X-ray clusters
($L_{X}> 5 \times 10^{44}$ erg s$^{-1}$), typically between redshifts of 0.2-0.4
\citep[e.g., the Extended Giant Meterwave Radio Telescope -GMRT- Radio Halo
Survey, EGRHS,][]{vent07,vent08,vent13,kale13,kale15}. However,
semi-analytical models and cosmological simulations have predicted that
sensitive low-frequency radio observations, such as those made with LOFAR at
$\sim 150$ MHz, should commonly find haloes in less massive systems
\citep{cass06,cass10a,cass12,zand14} as well as thousands of more radio relics
\citep{hoef11,nuza12}.

Abell 959 (hereafter A959), the subject of this study, is situated at a redshift
of $z=0.288$, has a mass of $M_{\rm SZ 500}= (5.08 \pm 0.47) \times 10^{15}$
M$_{\odot}$ \citep{plan13} and multiple galaxy concentrations \citep{bosc09}.
Multiple mass concentrations in A959 were identified from a weak gravitational
lensing analysis \citep{dahl02,dahl03}. Among these concentrations is a putative
dark mass clump (WL 1017.3+5931) that is not associated with a known galaxy
concentration or X-ray gas clump. Furthermore, \citet{bosc09}, using
spectroscopic observations, found a redshift of $z=0.288$, lower than the value
of $z=0.353$ used previously in the literature \citep[see also][]{irge02}. They
concluded that the cluster is in an early, dynamical stage of formation and
might be forming along two main directions of mass accretion. Diffuse radio
emission in A959 was reported in \citet{coor98} and \citet{owen99}, and the
latter found a flux density at 1.4 GHz of 3 mJy and a size of 0.8 Mpc. However,
A959 has not been studied at lower frequencies or in detail in X-rays up to now.

In this paper we present the results of a multiwavelength study of A959. We use
LOFAR data to study the radio emission and X-ray data from the XMM-\emph{Newton}
and \emph{Chandra} X-ray observatories to measure the cluster properties and to
place constraints on gas mass fraction of the putative dark mass clump (WL
1017.3+5931). Using a large sample drawn from the literature,
we place A959 in context with other RH systems and we investigate
the evolution of the X-ray and radio properties of RH clusters.
We adopt $H_{0}=70$ km s$^{-1}$Mpc$^{-1}$, $\Omega_{\Lambda}=0.7$,
and $\Omega_{\rm{M}}=0.3$ throughout.

\section{Data Analysis}\label{S:analysis}

\subsection{LOFAR Data}\label{S:radio_analysis}

A959 was observed with the High-Band Array (HBA) of LOFAR at frequencies of 120-170 MHz on 25-04-2015 for
8 hours as part of observing program LC3\_008 \citep[taken as
part of LoTSS, the LOFAR Two-meter Sky Survey;][]{shim17}. A 10-minute
observation of a calibrator, 3C196, was made immediately preceding the A959
observation and is used to set the overall flux \citep{scai12} and to remove
instrumental phase effects from the visibility data. Preprocessing of the data
from both observations  included flagging of radio-frequency interference (RFI) and
averaging in time and frequency (to reduce the raw visibility data to a
manageable size). These preprocessed data were obtained from the LOFAR long-term
archive and further processed using the \textsc{prefactor}\footnote{Available at
\url{https://github.com/lofar-astron/prefactor}} and \textsc{factor} pipelines
\footnote{Available at \url{https://github.com/lofar-astron/factor}}
to calibrate and image the data using the
facet-calibration scheme described in \citet{vanw16}. Version 2.0.2 of
\textsc{prefactor} and version 1.3 of \textsc{factor} were used.

The \textsc{prefactor} pipeline first derives the bandpass calibration and
corrects for instrumental phase effects using the 3C196 calibrator observation.
For each station, amplitude and phase corrections, plus an additional term that
tracks the rotation angle between the XX and YY phases, were solved for each
of the XX and YY polarizations every 4 seconds and 48.8 kHz. The model of 3C196
of XX was used for the calibration. For each time slot and station, the phase
solutions are then fit with a model that is comprised of a clock term that
scales with the frequency, $\nu$, a differential total electron content (dTEC)
term that scales as $\nu^{-1}$, and an offset term that is constant in
frequency. The clock and offset solutions are then transferred, along with the
amplitudes, to the target data. In this way, the direction-independent
instrumental effects are corrected for.

Next, the \textsc{prefactor} pipeline groups the data into bands of $\approx
2$~MHz each, the maximum bandwidth over which frequency-dependent effects can be
largely ignored (and therefore fit with a single solution in frequency). Each of these
bands is then phase calibrated using a model of the field obtained from the TIFR GMRT Sky Survey catalog \citep[TGSS,][]{inte17}
and imaged. The imaging is done in two passes, with the purpose of
modeling the sources in the field out to the second null of the primary beam. To
this end, two images are made of each band: one at a resolution of $\sim
25$\arcsec, used to detect and model the compact emission, and one at a
resolution of $\sim 75$\arcsec, used to model any diffuse, extended emission not
picked up in the higher-resolution image. The lower-resolution image is made of
the residual visibilities, after subtraction of the higher-resolution clean
components. Components from both images are then subtracted from the uv-data to
produce ``source-free'' datasets suitable for use in \textsc{factor}.

After \textsc{prefactor} was run, \textsc{factor} was used to correct for
direction-dependent effects. The main direction-dependent effects in HBA LOFAR
data are due to phase delays induced by the ionosphere and amplitude errors that
occur due to inaccuracies in the LOFAR beam model. \textsc{factor} corrects for
these effects by faceting the field and solving for a single set of corrections
for each facet. The field was divided into 45 facets, of which 12 were
processed. The processed facets were those that contained very bright sources
and those that neighbored on (or included) A959 (the 33 unprocessed
facets contain only fainter, more distant sources that do not affect the Abell
959 facet). \textsc{factor} was run with the default parameters. The full bandwidth
was used in the imaging, resulting in an image with a frequency of 143.7 MHz and an rms noise of
103~$\mu$Jy~beam$^{-1}$ at the field center.

The global flux scale was checked by extracting the LOFAR flux densities of the
41 brightest unresolved sources in the processed facets and comparing them to the
TGSS flux densities. We found the average ratio of LOFAR-to-TGSS flux density to
be 1.05, approximately the ratio expected given the slightly different frequencies of
the images (143.7 MHz for LOFAR and 150 MHz for the TGSS) and the average spectral
index of radio sources ($\approx -0.8$). We adopt a conservative systematic
uncertainty of 15\% on all LOFAR flux densities throughout our analysis, as done in previous LOFAR-HBA works.

\begin{figure*} \begin{tabular}{@{}cc}
\includegraphics[width=84mm]{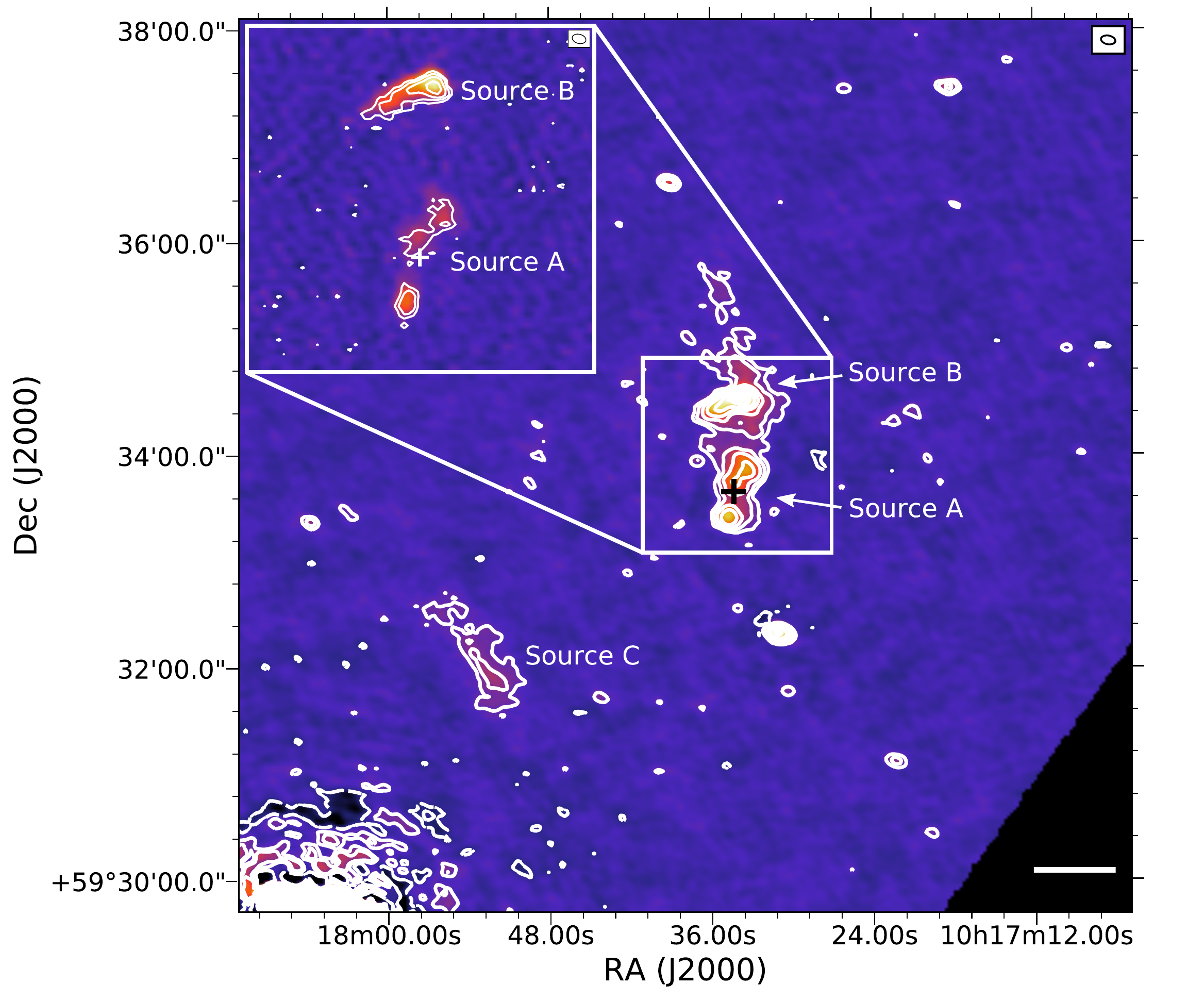} &
\includegraphics[width=84mm]{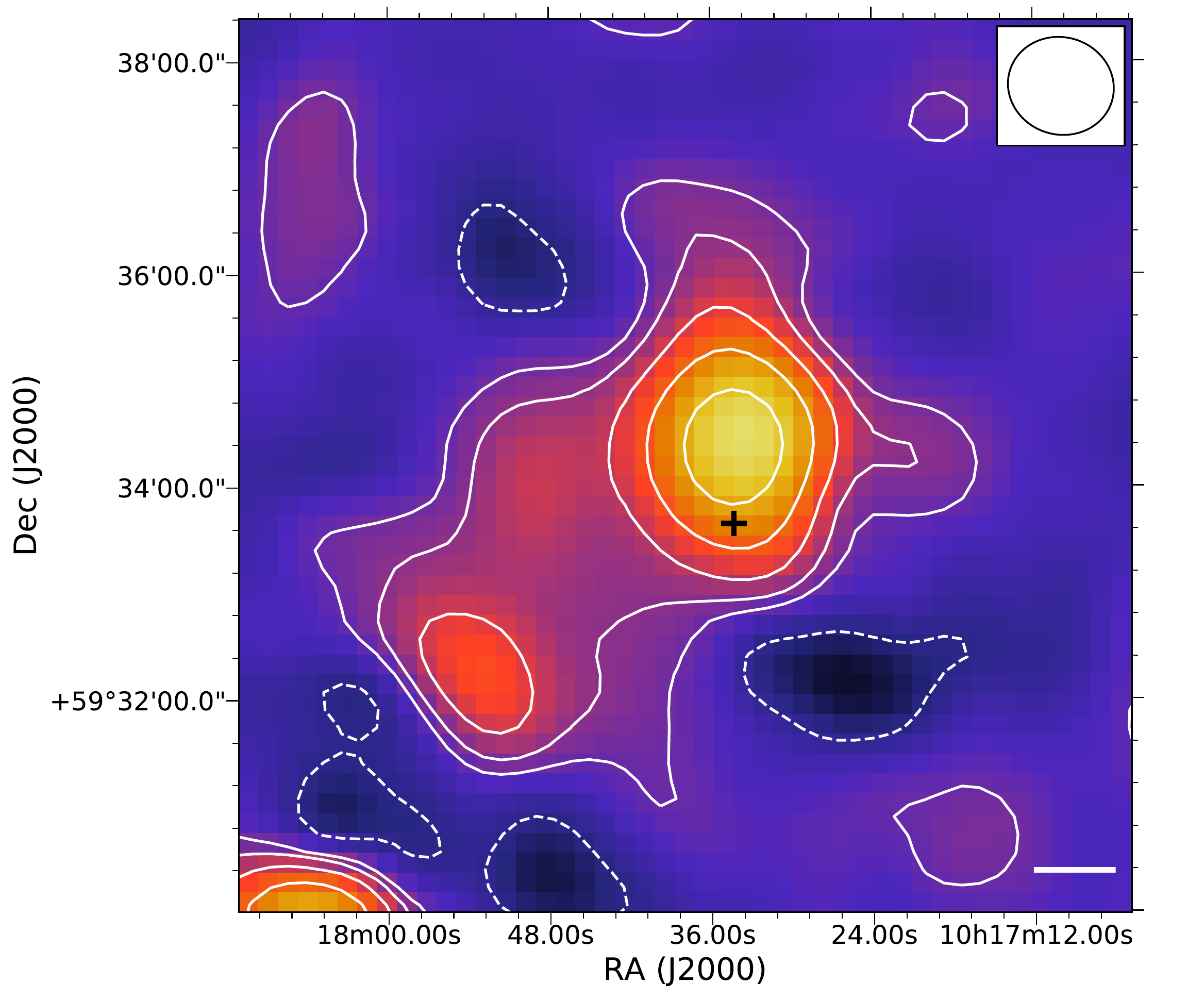} \\
\includegraphics[width=84mm]{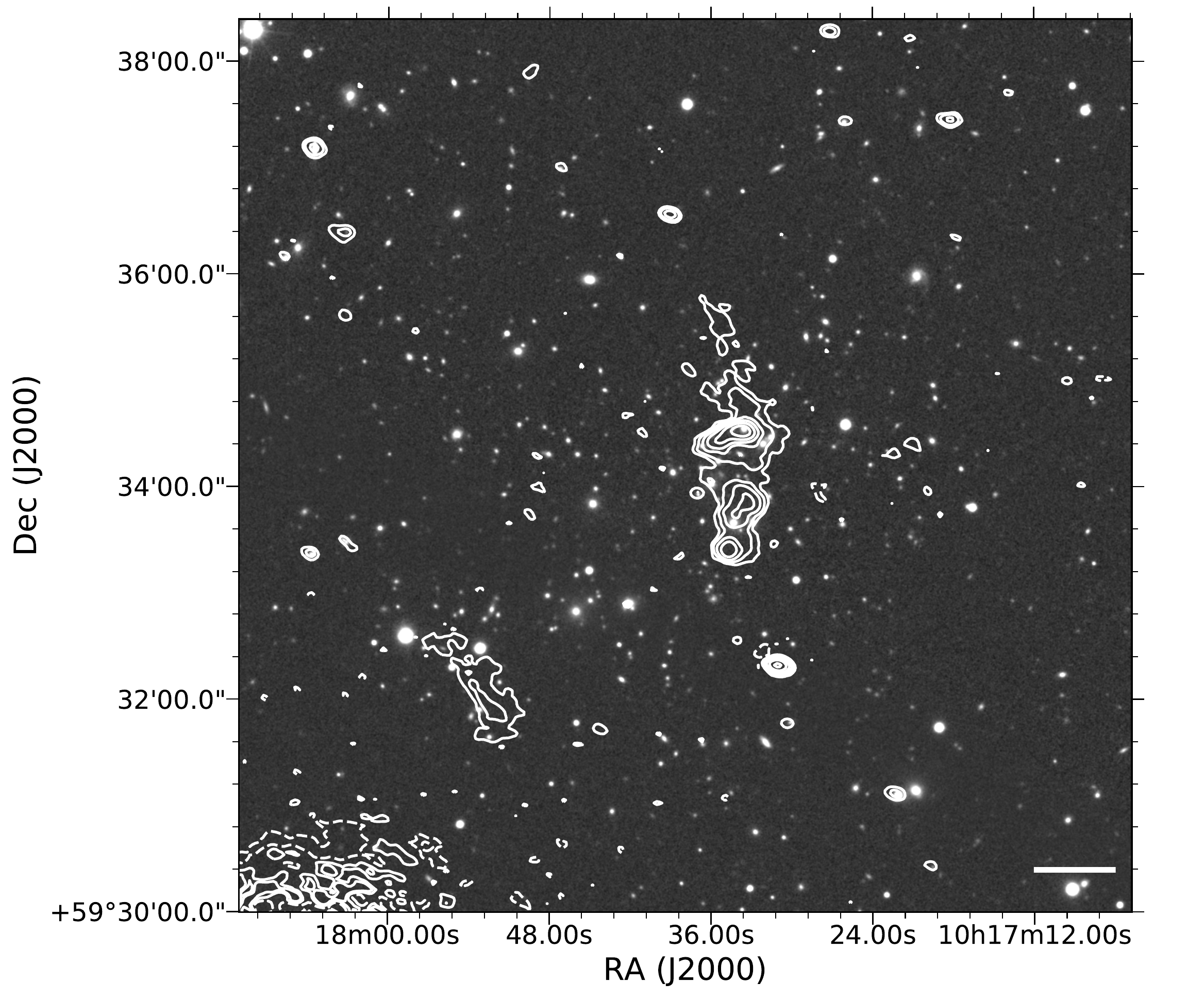} &
\includegraphics[width=84mm]{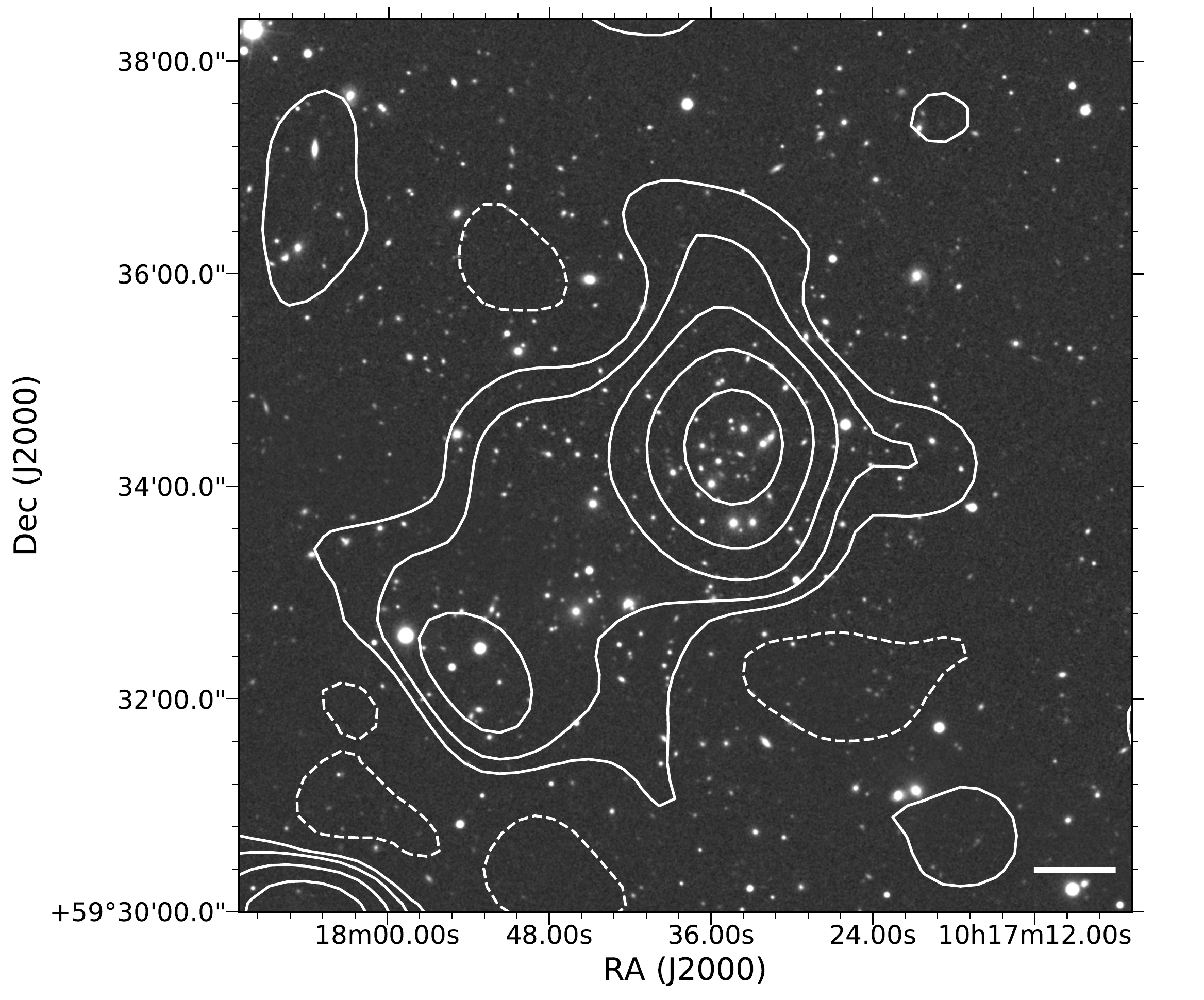} \\
\end{tabular}
\caption{\emph{Top row:} LOFAR images at 143.7 MHz at high-resolution (\emph{left}) and
low-resolution (\emph{right}) after subtraction of the compact emission. The compact
emission that was subtracted is shown in the inset image in the left panel (see text for
details). The restoring beam is indicated by the white ellipse
in the upper right-hand corner, and the scale bar represents 200 kpc at the redshift
of A959. Contours begin at 3 times the rms noise of 118.5 $\mu$Jy beam$^{-1}$ and 426.7 $\mu$Jy
beam$^{-1}$ for the high- and low-resolutions images, respectively, and increase by a factor
of 2. The first negative contour (at -3 times the rms noise) is also plotted and is denoted by the dashed lines.
The cross marks the location of the BCG. \emph{Bottom row:} SDSS optical $r$-band
image with the contours from the high-resolution (\emph{left}) and
low-resolution (\emph{right}) LOFAR images overlaid. In top-\emph{left} panel: Source A
is the central radio source associated with the BCG (see Section \ref{S:radio_agn});
source B is a tailed radio galaxy and has a flux $S_{143.7{\rm ~MHz}} = 45.3$~mJy,
and Source C is the candidate radio relic (see Section \ref{S:radio_relic}).}
\label{F:lofar_images}
\end{figure*}

Figure~\ref{F:lofar_images} shows two images at 143.7~MHz: a high-resolution
image, with a restoring beam with a FWHM of $4.9\arcsec \times 8.3$\arcsec, and a
low-resolution residual image, made after subtracting compact emission,
with a restoring beam of $55\arcsec \times 60$\arcsec. The compact emission was modeled by imaging
with a uv minimum of 4 k$\lambda$, a cut that results in emission on scales of $\gtrsim 60$ arcsec
being excluded (see Figure~\ref{F:lofar_images}). The resulting clean components were then subtracted from the
visibilities (using the \emph{ft}
and \emph{uvsub} tasks in CASA v4.7.1) and the low-resolution
residual image made by tapering the uv-data with a Gaussian taper to achieve a
resolution of $\approx 40$ arcsec.

In the full-resolution image, a number of features are apparent: a source
(source A) that is associated with the brightest cluster galaxy (BCG), with two
lobes oriented approximately N-S; a source (source B) that is located to the north of the BCG
and appears to be a head-tail radio galaxy; and a linear, relic-like feature
(Source C) that does not appear to be clearly associated with any optical galaxy.
In the low-resolution residual image, diffuse emission is seen that fills most
of the region between the BCG and the relic-like source C.
We will discuss these features in detail in Section~\ref{S:results}.

\subsection{GMRT Data}\label{S:gmrt_analysis}

GMRT 325~MHz observations of A959 were obtained on 06-03-2017 (project ID
31\_009; PI de Gasperin). Visibilities were recorded over 33.3~MHz of bandwidth,
starting with 20~minutes on calibrator 3C147, then 213~minutes on A959, and
finally 16~minutes on 3C147 again. The data were processed using the SPAM
pipeline \citep{inte17} in the default mode, and calibrated using 3C147 while
adopting the flux scale from \citet{scai12}. This resulted in a final image with
a central frequency of 322.7~MHz and an rms noise of 84~$\mu$Jy beam$^{-1}$ at
the field center.

\begin{figure}
\includegraphics[width=84mm]{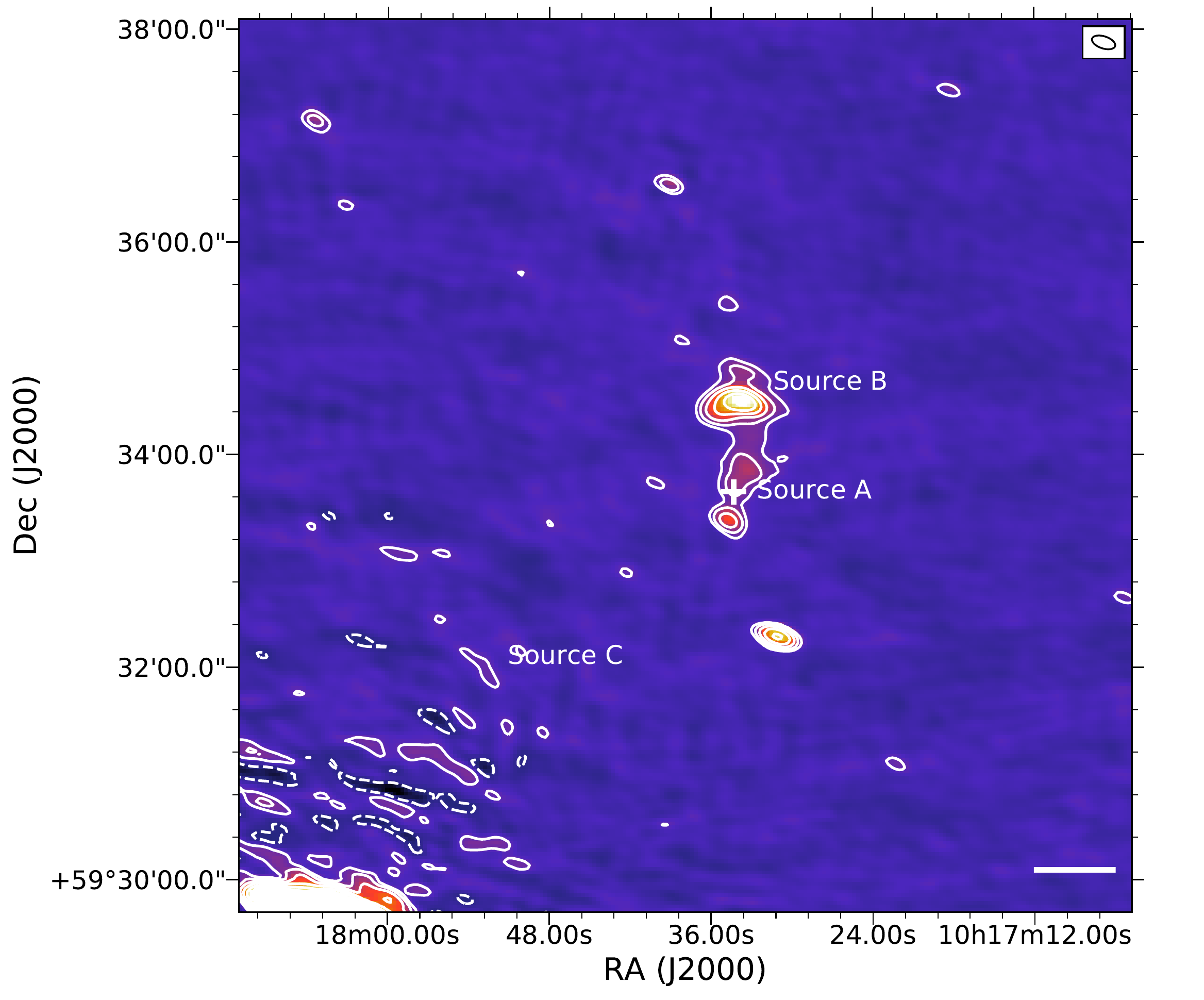}
\caption{GMRT image at 322.7 MHz. Contours begin at 3 times the rms noise of
130.6 $\mu$Jy beam$^{-1}$. The restoring beam is $6.9\arcsec \times
13.7$\arcsec. The scale bar, symbols, and annotations are the same as in
Figure~\ref{F:lofar_images}, left.}
\label{F:gmrt_images}
\end{figure}

The resulting 322.7~MHz GMRT image is shown in Figure~\ref{F:gmrt_images}, with
the sources identified in the high-resolution LOFAR image labeled. As with the
LOFAR data, we searched for diffuse emission by modeling and subtracting the
compact emission and imaging the residual data at lower resolution, but we did
not detect any such emission. However, sources A and B are clearly detected in
the GMRT image with very similar morphologies to those in the LOFAR image.
Source C, the putative relic, is not detected (there is a hint of emission at
its location, but its significance is low and may be a sidelobe of the
bright source nearby).

\subsection{\emph{Chandra} Data}\label{S:chandra_analysis}

A959 was observed by the \emph{Chandra} X-ray Observatory on 01-02-2016 for
7.6 ks (ObsID 17161, VFAINT mode) with the \mbox{ACIS-I} instrument. The data
were obtained from the \emph{Chandra} data archive and were reprocessed with
\textsc{ciao} 4.8\footnote{See \url{cxc.harvard.edu/ciao/index.html}.} using
\textsc{caldb} 4.7.3\footnote{See \url{cxc.harvard.edu/caldb/index.html}.}. The
data were corrected for known time-dependent gain and charge transfer
inefficiency problems, and the events files were filtered for flares using the
\textsc{ciao} script \emph{lc\_clean} to match the filtering used during the
construction of the blank-sky background files used for background
subtraction.\footnote{See \url{http://asc.harvard.edu/contrib/maxim/acisbg/}.} A
total of 7.1 ks remained after filtering. The background file was normalized to
the count rate of the source image in the $10-12$ keV band (after filtering).
Lastly, point sources detected using the \textsc{ciao} tool \emph{wavdetect}
were removed.

Spectra were extracted in annuli constructed to contain at least 500 counts each
using the \textsc{ciao} script \emph{specextract}. For each spectrum, weighted
responses were made, and a background spectrum was extracted in the same region
of the CCD from the associated blank-sky background file. For the spectral
fitting, \textsc{xspec} \citep{arna96} version 12.7.1 was used. Gas temperatures
and densities were found by deprojecting these spectra using the Direct Spectral
Deprojection method of \citet{sand07}. The deprojected spectrum
in each annulus was then fit in \textsc{xspec} with a single-temperature plasma
model (MEKAL) absorbed by foreground absorption model (WABS), between the
energies of 0.5 keV and 7.0 keV. In this fitting, the redshift was fixed to
$z=0.288$ \citep{bosc09}, and the foreground hydrogen column density was fixed to $N_{\rm H} =
8.78 \times 10^{19}$~cm$^{-2}$, the weighted-average Galactic value from
\citet{dick90}.

The density was calculated from the normalization of the MEKAL component,
assuming $n_{\rm{e}}=1.2n_{\rm{H}}$ (for a fully ionized gas with hydrogen and
helium mass fractions of $X=0.7$ and $Y=0.28$).  The pressure in each annulus
was calculated as $P=nkT,$ where we have assumed an ideal gas taking $n =
2n_{\rm{e}}$. The entropy is taken as $S=kTn_{\rm e}^{-2/3}$. The cooling time
was derived from the temperature, metallicity, and density using the cooling
curves of \citet{smit01}.

We also derived the X-ray luminosity and emission-weighted
temperature inside the $R_{500}$ region, defined as the region at which the
mean mass density is 500 times the critical density at the cluster
redshift \citep[see][]{prat09} \footnote{$R_{500}=(\frac{M_{500}}{500
\rho_c(z)  4 \pi/3})^{1/3}$, with $\rho(z)=\frac{E(z)^{2} 3 H_{0}^{2}}{8
\pi G}$ and $E(z)^{2}=\Omega_{\rm{M}}(1+z)^{3}+\Omega_{\Lambda}$}. We
found $R_{500}=1100$ kpc using the mass $M_{500}=(5.08 \pm 0.47) \times$
10$^{14}$ M$_{\odot}$, derived from the SZ signal $Y_{\rm{SZ}}$
\citep{plan15}.

We fit a spectrum extracted from this region between 0.5-7.0 keV in XSPEC
(model \emph{wabs*cflux*mekal}) with the abundance
fixed at $Z=0.3$ $Z_{\odot}$ \citep{mern17}.
We found a global temperature of $kT=6.05 \pm 1.13$ keV and an X-ray luminosity within
$R_{500}$ of $L_{\rm X 500}=(4.51 \pm 0.33) \times$ 10$^{44}$
erg s$^{-1}$ in the 0.5-7.0 keV band and $L_{\rm X 500}=(2.36 \pm 0.17)
\times$ 10$^{44}$ erg s$^{-1}$ in 0.5-2.4 keV band. In the 0.1-2.4 keV
band (the ROSAT X-ray band), we found a X-ray luminosity within $R_{500}$
of $L_{\rm X 500}=(2.77 \pm 0.18) \times$ 10$^{44}$ erg s$^{-1}$.

\subsection{XMM-\emph{Newton} Data}\label{S:xmm_analysis}

\begin{figure*} \begin{tabular}{@{}cc}
\includegraphics[width=84mm]{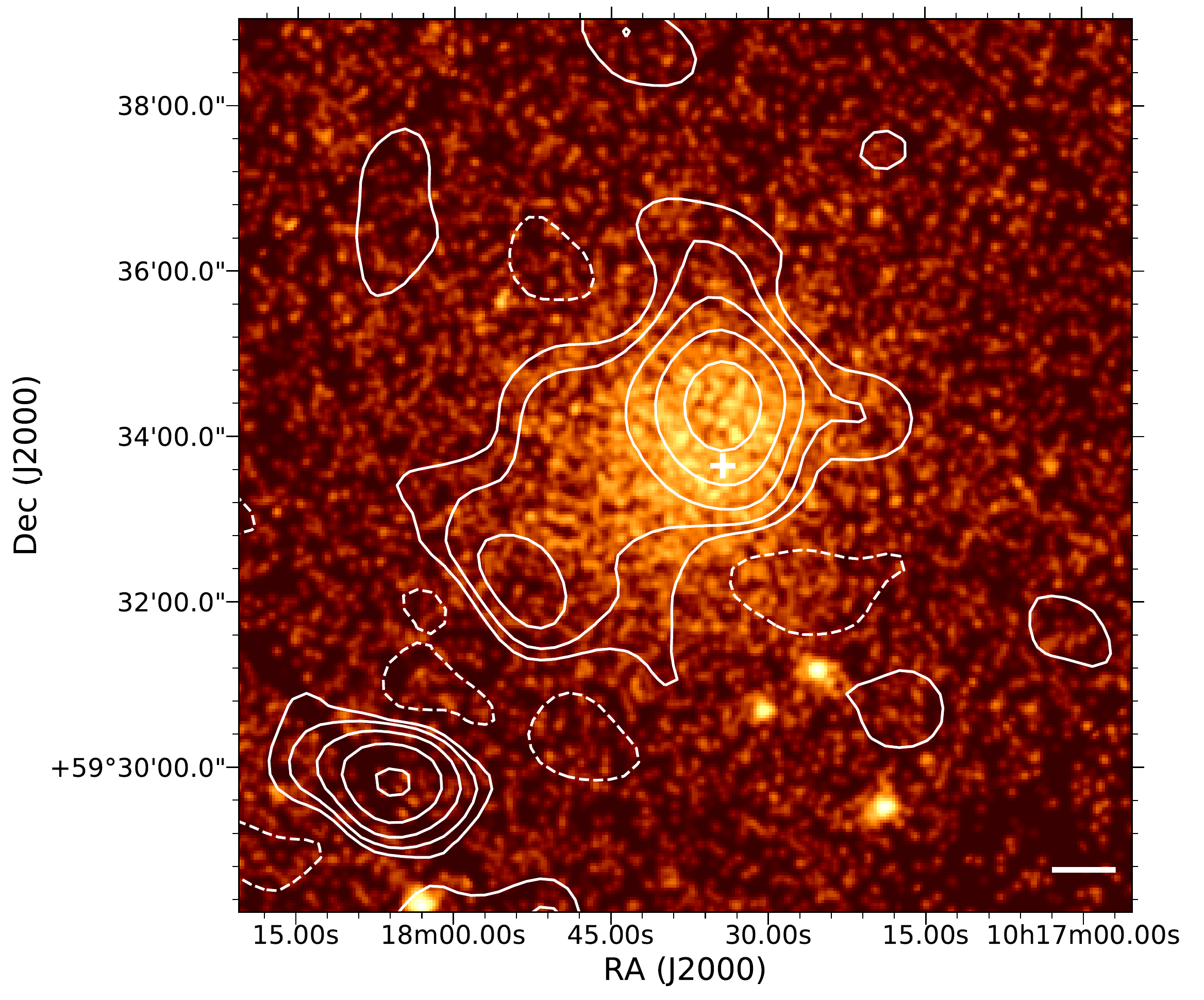} &
\includegraphics[width=84mm]{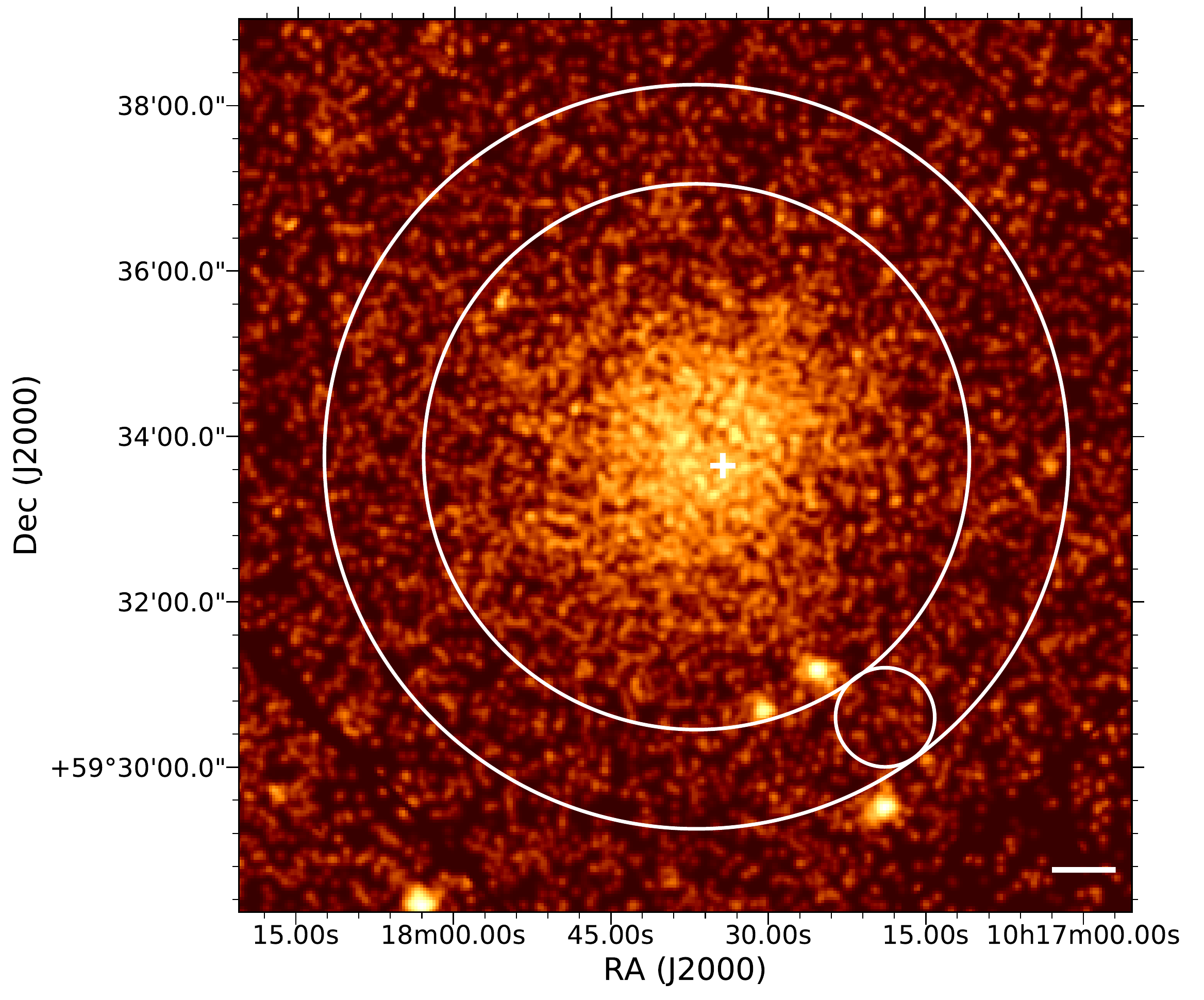} \\
\end{tabular}
\caption{Combined MOS+PN XMM-\emph{Newton} X-ray image, with the contours from the low-resolution
residual LOFAR image (\emph{left}) and the regions used in the dark-clump analysis (\emph{right}) overlaid.
The image has been smoothed by a Gaussian with FWHM = 5 pixels. The scale bars and symbols are the same as in
Figure~\ref{F:lofar_images}.}
\label{F:xray_images}
\end{figure*}

A959 was observed by the XMM-\emph{Newton} X-ray Observatory on 12-04-2007
for 41.5~ks (Obs.~ID 0406630201). The data were obtained from the
XMM-\emph{Newton} archive and were processed with the \emph{epchain} and
\emph{emchain} tasks in \textsc{xmmsas} version 16.0.0. Periods of background
flaring were identified as times for which the total count rate exceeded 0.35
and 0.4~count~s$^{-1}$ for the MOS and PN detectors, respectively.
Unfortunately, $\sim$ 90\% of the data was affected by a strong flare: after filtering
periods of high background, only 9.785 ks for the MOS detectors and 4.999 ks for
the PN detector remained.

Exposure-corrected images were made with the \emph{evselect} and
\emph{eexpmap} tasks from the cleaned event lists between the energies of
0.5-2.5~keV, where the signal-to-noise of the soft thermal cluster emission is
greatest. These images were then used to constrain the emissivity of the dark
clump to the south of the main cluster (see Section~\ref{S:dark_clump}). The
background in the region of the dark clump is dominated by the cluster emission.
For this region, we use as the background count rate the mean
count rate in an annulus, centered on the cluster, with inner and outer radii
that match those of the dark clump (see Figure~\ref{F:xray_images}).

We also extracted a spectrum within the $R_{500}$ region ($R_{500}=1099$ kpc)
using the MOS1 data and a local background region that is free of any cluster emission. We
fitted the above spectrum in XSPEC with a fixed $N_{\rm H}$, fixed redshift and
fixed abundance $Z=0.3$ $Z_{\odot}$, and found a temperature of $kT=8.55 \pm
2.30$ keV and a X-ray luminosity within $R_{500}$ region of  $L_{\rm X
500}=(4.97 \pm 0.35) \times$ 10$^{44}$ erg s$^{-1}$ in the 0.5-7.0 keV band and
$L_{\rm X 500}=(3.24 \pm 0.46) \times$ 10$^{44}$ erg s$^{-1}$ in the 0.1-2.4 keV
band. Therefore, the \emph{Chandra} and XMM-\emph{Newton} values for luminosity
and temperature agree within the 1-$\sigma$ errors. For convenience, we will use
the luminosity derived from the XMM-\emph{Newton} data in further calculations.

\section{Results and Discussion}\label{S:results}

\subsection{X-ray Properties}\label{S:xray_props}

The appearance of the ICM of A959 is fairly smooth, with no cuspy core or
other bright substructures (excluding the X-ray point sources, see
Figure~\ref{F:xray_images}). A number of faint, associated galaxy
groups (or subclusters) have been identified previously in ROSAT observations
\citep[see,][]{dahl03,bosc09}.

The spectral analysis of the X-ray data (see Sections~\ref{S:chandra_analysis} and \ref{S:xmm_analysis})
indicates that the temperature of the ICM is $kT \approx 6$-7 keV. There is no evidence of
cooler gas in the core. The central density is $n_e \approx 2 \times
10^{-3}$~cm$^{-2}$ and the central cooling time is $t_{\rm cool} \approx 3 \times
10^{10}$~yr. A959 is therefore a typical massive NCF cluster. It
shows no evidence for possessing a cool corona associated with the BCG, as seen in some
NCFs such as the Coma cluster \citep{sun09}.

The X-ray luminosity within the $R_{500}$ region in the 0.1-2.4 keV band derived
using \emph{Chandra} and XMM-\emph{Newton} data (see Section
\ref{S:chandra_analysis} Section \ref{S:xmm_analysis}) is a factor of three less
than the MCXC value from \citet{piff11} of $L_{\rm X 500}=8.37 \times$ 10$^{44}$
erg s$^{-1}$ (after correcting for the revised redshift). This factor of three
is too large to be due only to the difference in $R_{500}$ used in MCXC catalog
\citep[$R_{500}=1260$ kpc after correcting for redshift,][]{piff11}. Instead,
the difference might be a result of uncertainties in the modeling that was used
for the ROSAT data to correct from aperture flux to $L_{\rm X 500}$. In support
of this possibility, we find a bolometric luminosity within $R_{500}$ from the
XMM-\emph{Newton} data of $L_{\rm X, bol, 500} = 7.1 \times$ 10$^{44}$ erg
s$^{-1}$, similar to the value derived by \citet{mahd13}, using the same
XMM-\emph{Newton} data, of $L_{\rm X, bol, 500}=5.8 \times$ 10$^{44}$ erg
s$^{-1}$, a difference of only $\approx 20$\% \citep[see also][]{conn14}.

\subsection{The Central Radio Source}\label{S:radio_agn}

The full-resolution LOFAR image, shown in Figure~\ref{F:lofar_images}, reveals
that the BCG in the cluster core is a radio galaxy with a bright core (centered
on the BCG) and lobes that extend $\sim 100$ kpc to the north and south. The
total flux density at 143.7~MHz, measured from the high-resolution LOFAR image,
is $S_{143.7{\rm ~MHz}} = 22.9 \pm 2.3$~mJy, corresponding to a luminosity of
$P_{143.7{\rm ~MHz}} = (5.8 \pm 0.6) \times 10^{24}$~W~Hz$^{-1}$. The source is
also detected in the 322.7~MHz GMRT image, with a flux density of $S_{322.7{\rm
~MHz}} = 9.1 \pm 1.1$~mJy, implying a spectral index ($\alpha$, where $S_\nu
\propto \nu^{-\alpha}$) between 143.7 and 322.7 MHz of $\alpha_{143}^{322} = 1.14
\pm 0.25$.

Adopting a power-law spectrum with this spectral index, we find a luminosity for
the central radio source at 1.4 GHz of $P_{1400{\rm ~MHz}} = (4.3 \pm 0.5)
\times 10^{23}$~W~Hz$^{-1}$. This luminosity is above the value of the threshold
between NCF and CF clusters seen in the B55 and HIFLUGCS cluster samples
\citep{birz12}.

The interaction between the lobes of the central radio source and the ICM should
create X-ray cavities which will rise buoyantly into the ICM. As they are
inflated and evolve, they do work on the surrounding ICM. This work is one
component of AGN feedback, the \emph{maintenance} or \emph{radio-mode} AGN
feedback \citep[for reviews see][]{mcna07,fabi12}. Such feedback is rare in NCF
systems, although evidence for cavities in a NCF system was recently found in
the SPT sample \citep[e.g., SPT-CL J2031-4037,][]{birz17}. Therefore, the
prevalence and importance of AGN feedback in NCF systems is not well
established, but there might often be radio activity and AGN feedback at a low
level in such systems. Deeper \emph{Chandra} data are required to identify any
cavities in the ICM of A959.

\subsection{The Giant Radio Halo}\label{S:radio_halo}

There is clear evidence for diffuse emission to the east of the X-ray core in
the low-resolution LOFAR image, shown in Figure~\ref{F:lofar_images}. This
emission extends from the central BCG to the relic, with a largest linear size
of $\sim 5$~arcmin=1.3~Mpc, although it does not uniformly fill this region. The
total flux density of the halo, excluding the candidate relic (see
Section~\ref{S:radio_relic}) and the compact emission from the BCG and the
head-tail source to the north of the BCG, is $S_{143{\rm ~MHz}} = 94 \pm
14$~mJy, corresponding to a luminosity of $P_{143{\rm ~MHz}} = (2.08 \pm 0.32)
\times 10^{25}$~W~Hz$^{-1}$. Using the 1400 MHz flux density of the diffuse
emission measured by \citet{owen99} of $S_{1400{\rm ~MHz}} = 3 \times
10^{-3}$~Jy, we find a luminosity of $P_{1.4{\rm ~GHz}} = 0.68 \times
10^{24}$~W~Hz$^{-1}$ and a spectral index of $\alpha_{143}^{1400} =
1.48^{+0.06}_{-0.23}$, where the error includes an estimate for the error in the
subtraction, adopted to be 50\% of the subtracted flux.\footnote{This spectral
index is consistent with the lack of a detection in the residual 322.7~MHz GMRT
image, given the noise in this image and the expected flux density of the halo
at 322.7 MHz.} The halo in A959 has a somewhat steeper spectrum than that of the
average giant radio halo \citep[$<\alpha> \approx 1.3$,][]{cass13}, but we note
that our value of $\alpha_{143}^{1400}$ should be treated with caution, as we do
not know exactly how the 1400 MHz image of \citet{owen99} differs in sensitivity
to diffuse emission from our 143 MHz image (e.g., due to different sampling of
the uv plane). Also, we do not know if any embedded discrete sources in the 1.4
GHz image were completely subtracted or whether the regions used for the
flux-density measurement are identical.

Diffuse radio emission in the form of a giant RH is often interpreted as
evidence of recent, energetic merging activity \citep{cass13}. Such activity is
expected in higher-redshift systems of X-ray flux limited samples \citep[e.g.;
GRHS, EGRHS,][]{vent07,kale15}, such as the one to which A959 belongs
\citep[see NORAS,][]{bohr00}. Below, we compare A959 with other systems which
posses giant radio haloes.

\subsection{Scaling Relations for Radio Haloes}\label{S:radio_halo_relations}

To date, there are approximately 80 systems with detected radio haloes \citep{fere12,vanW19}. For these
systems, the radio luminosity of the RH is known to scale with a number of cluster
properties, the most commonly used of which are the cluster mass, the cluster SZ
signal ($Y_{\rm SZ}$), and the X-ray luminosity \citep[see][]{cass13,mart16}.\footnote{X-ray
luminosity, cluster mass and SZ signal are calculated within $R_{500}$, and the
X-ray luminosity is measured in the 0.1-2.4 keV band.}
These relations were derived using a sample of $\approx$ 25 systems
in \citet{cass13} and 41 systems in \citet{mart16} drawn from the
literature, 11 of which are from the GRHS/EGRHS sample
\citep[see][]{vent07,vent08,vent13,kale13,kale15}. These RH samples are
comprised of systems with a wide range of redshift, mostly between $0.05<z<0.55$,
with the notable exceptions of Coma at $z=0.023$ and El Gordo at $z=0.87$.

Additionally, there are a number of RHs known from other studies of single systems and smaller
samples that are not present in the above samples \citep[e.g., A399,
A401, A2218, A2061, A2065, A2069, PLCKG287.0+32.9, MACS J0416.1-2403
etc;][]{fere12,giov00,rudn09,farn13,bona14,ogre15}. Also, some systems from the GRHS
or EGRHS are not present in the above samples  (e.g.; A1682, A2261,
RXCJ1314.4-2515, ZwCL5247).   Lastly, in the last two years, there
has been a rapid increase in studies of individual or small samples of RH
systems
\citep[see][]{bern16,know16,gira16,vent17,pare17,duch17,hoag17,wilb18,hlav18,
cuci18,savi19}. We have collected measurements from these studies to form a
larger sample of RHs. The systems that are not present in \citet{cass13} or
\citet{mart16} samples are listed in Table \ref{RH_table}, with nine of these also
present in the \citet{yuan15} sample (A399, A2061, A2069, A2218, A3562,
ZwCL5247, CL0217+70, H1821+643, and the ''Toothbrush'' cluster). Table
\ref{Lx_table} lists the X-ray luminosities; for systems in the \citet{cass13} and
\citet{mart16} samples the cluster masses and the RH powers are listed in the
above papers.

We note that the halo powers in this larger sample have not been
derived in a homogeneous way. For example, in some cases the contribution of
compact radio sources could not be fully isolated from the RH emission
\citep[e.g., A2065 and A2069;][]{farn13}\footnote{We did not include
A2390 from \citet{somm17}, as it was not confirmed by LOFAR observations
\citet{savi19}, and A1914 and A2146 since they have only putative RH emission in recent LOFAR observations
\citep{mand19,hoag19}.}.
Also, the cluster X-ray luminosities were not derived in a homogenous way: we
used the values from \citet{cass13} where available, otherwise we used other
samples with derived X-ray luminosities
\citep[e.g.;][]{ohar06,mant10,gile17,yuan15}, or individual papers in some cases
when available \citep[e.g.; A1132, ACT-CL J0256.5+0006, CIZA
J2242.8+5301;][]{wilb18,know16,hoag17}\footnote{For PLZ1G139.61+24.20,
PLZ1G108.18-11.53 and ZwCL2341.1+0000  we derived the X-ray luminosity using the
archived \emph{Chandra} data (Obs IDs=15139, 17312, 17490).}. Otherwise, we used
the values from \citet{piff11} and even bolometric X-ray luminosity in some
cases (e.g.; CL1446+26)\footnote{Some systems are present in more than one of
the above studies \citep{ohar06,piff11,cass13,mant10,gile17}. In general, the
X-ray luminosities between studies are consistent, with a few exceptions where
there is a factor of 2 or more difference between studies e.g., A2142, A2261,
A141, and A1689.}. Due to these inhomogeneities, we do not attempt to derive new
scaling relations; rather, our goal here is to collect a sample of RHs in order
to search for more general trends.

In Figure \ref{F:Phalo_vs_L_M}, we plot the halo radio power versus the cluster X-ray
luminosity between 0.1-2.4 keV within R$_{500}$ (see Table
\ref{RH_table}) and cluster mass within R$_{500}$ derived from SZ observations
(see Table \ref{Lx_table}) for the larger sample of 80 systems
described above (A959 plus the literature systems). However, some systems in this sample do not
have X-ray luminosities available in the literature (e.g., PSZ1G018.75+23.57), and hence they do
not appear in the right panel of Figure \ref{F:Phalo_vs_L_M}. Additionally,
others do not have SZ-derived masses available in the literature (e.g., A523, A800, A851, MACS
J0416.1-2403, CIZA J2243.8+5301, CL0217+70, CL1446+26) and do thus not appear in the
left panel of Figure \ref{F:Phalo_vs_L_M}.

Figure \ref{F:Phalo_vs_L_M} shows that there is a large scatter about the above
scaling relations \citep[for a discussion see][]{brun09,basu12,cass13,cuci18}.
Some of the scatter in the radio power versus X-ray luminosity plot is likely intrinsic,
due to for example different systems being caught in different stages of the merger.
Significant changes in the X-ray luminosity are expected to occur during and after the merger event
\citep[e.g.;][]{rick01,ritc02,rand02,donn13}. In addition, the radio properties of RHs
are predicted to depend on the details of the merger (e.g., mass ratio and energetics) and will evolve
during the merger, thus introducing additional scatter \citep[see][]{cass13,mart16,cuci18}.

\subsection{The Relation of Cluster Properties to the Merger State}\label{S:lrad_vs_ratio_rel}

\begin{figure*} \begin{tabular}{@{}cc}
\includegraphics[width=84mm]{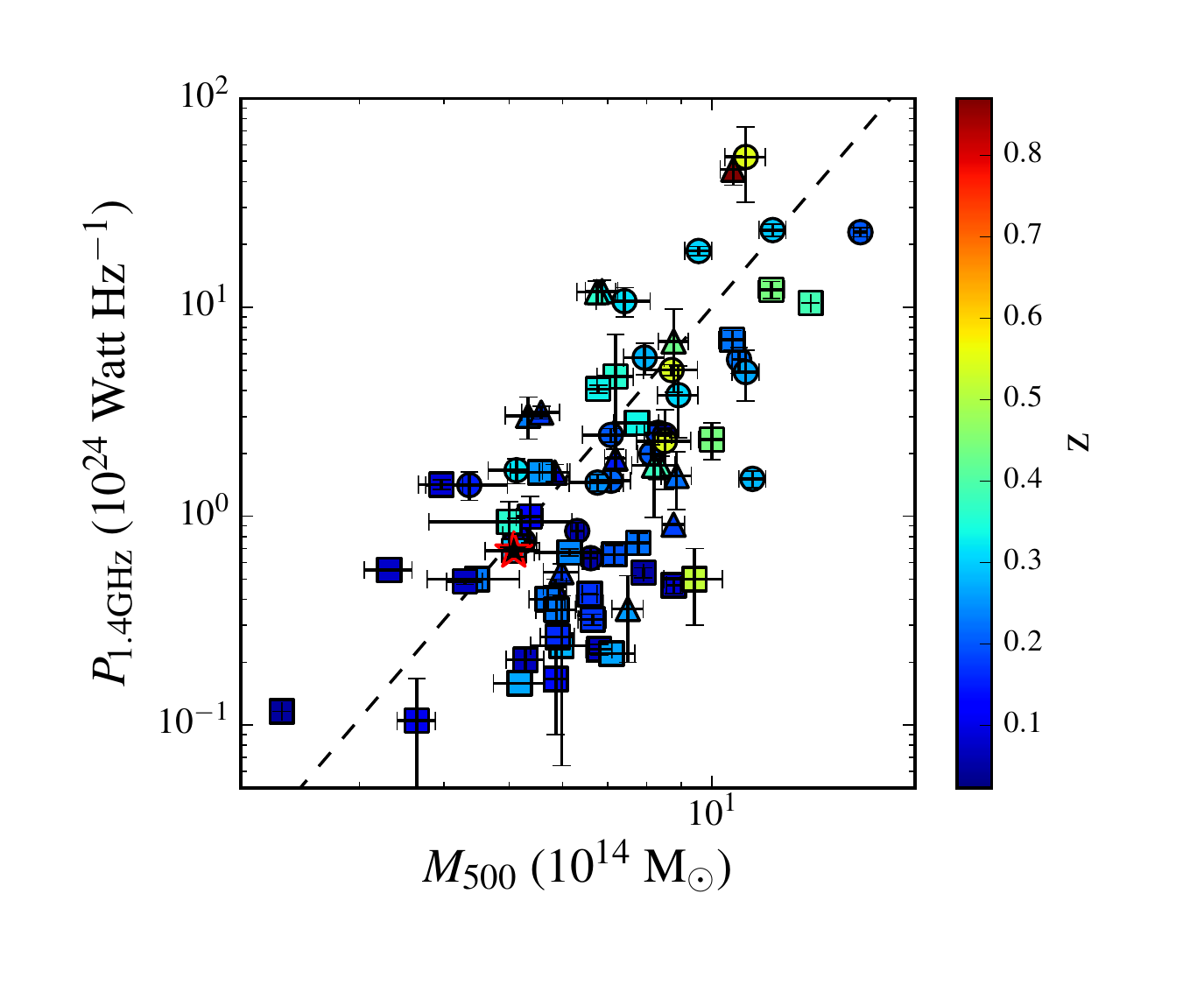} &
\includegraphics[width=84mm]{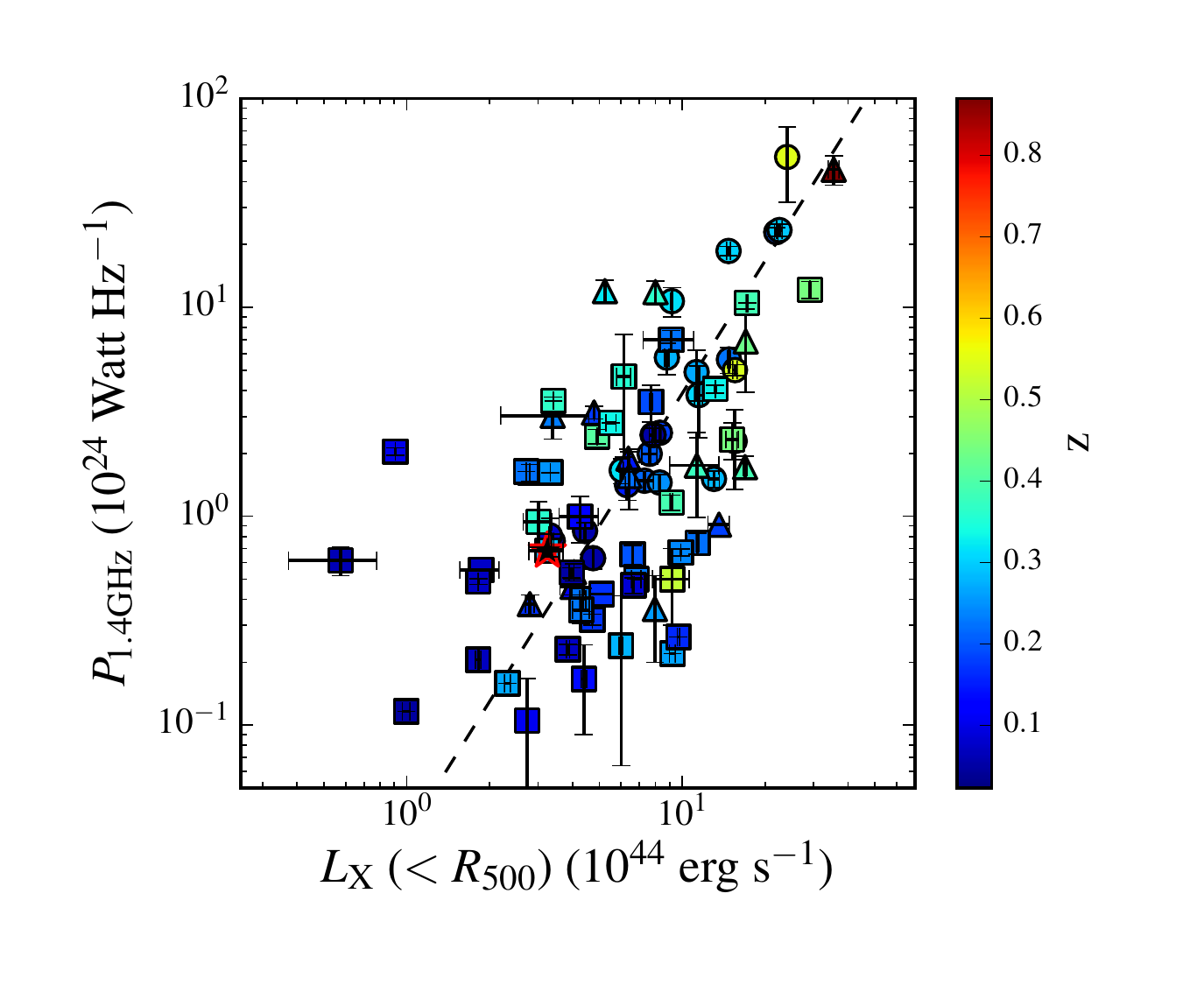} \\ \end{tabular}
\caption{The monochromatic 1.4 GHz radio power versus the SZ-derived cluster mass
$M_{500}$ (\emph{left} panel) and versus the X-ray luminosity in the 0.1-2.4 keV band
$L_{\rm X}(<R_{500})$ (\emph{right} panel), both derived within $R_{500}$. Except for A959
(denoted by the red star), the values for the systems are taken from the
literature (see Table 1 and Table 2). However, for El Gordo we used the radio halo power from \citet{lind14}. Circles denote the
systems from the \citet{cass13} sample, triangles denote the extra 16 systems from \citet{mart16} sample and the squares denote the systems from Table \ref{RH_table}. The
dashed lines show the best-fit relations of \citet{cass13}. Some systems do not appear in
the left panel since there are no available SZ-derived masses, while others do not appear in the right panel since there are no published X-ray luminosities (e.g; PSZ1G018.75+23.57).}
\label{F:Phalo_vs_L_M}
\end{figure*}

\begin{figure*} \begin{tabular}{@{}cc}
\includegraphics[width=84mm]{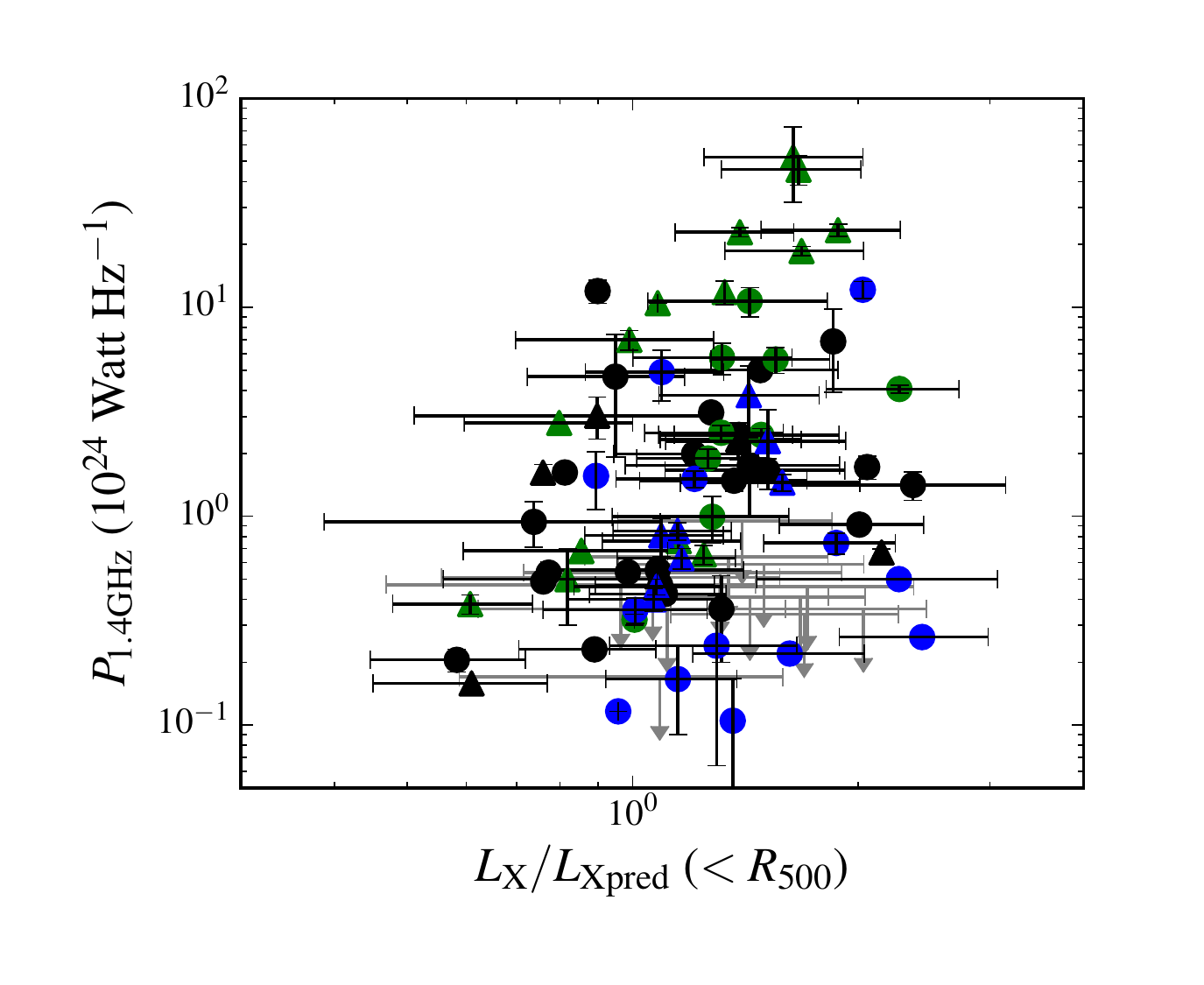} &
\includegraphics[width=84mm]{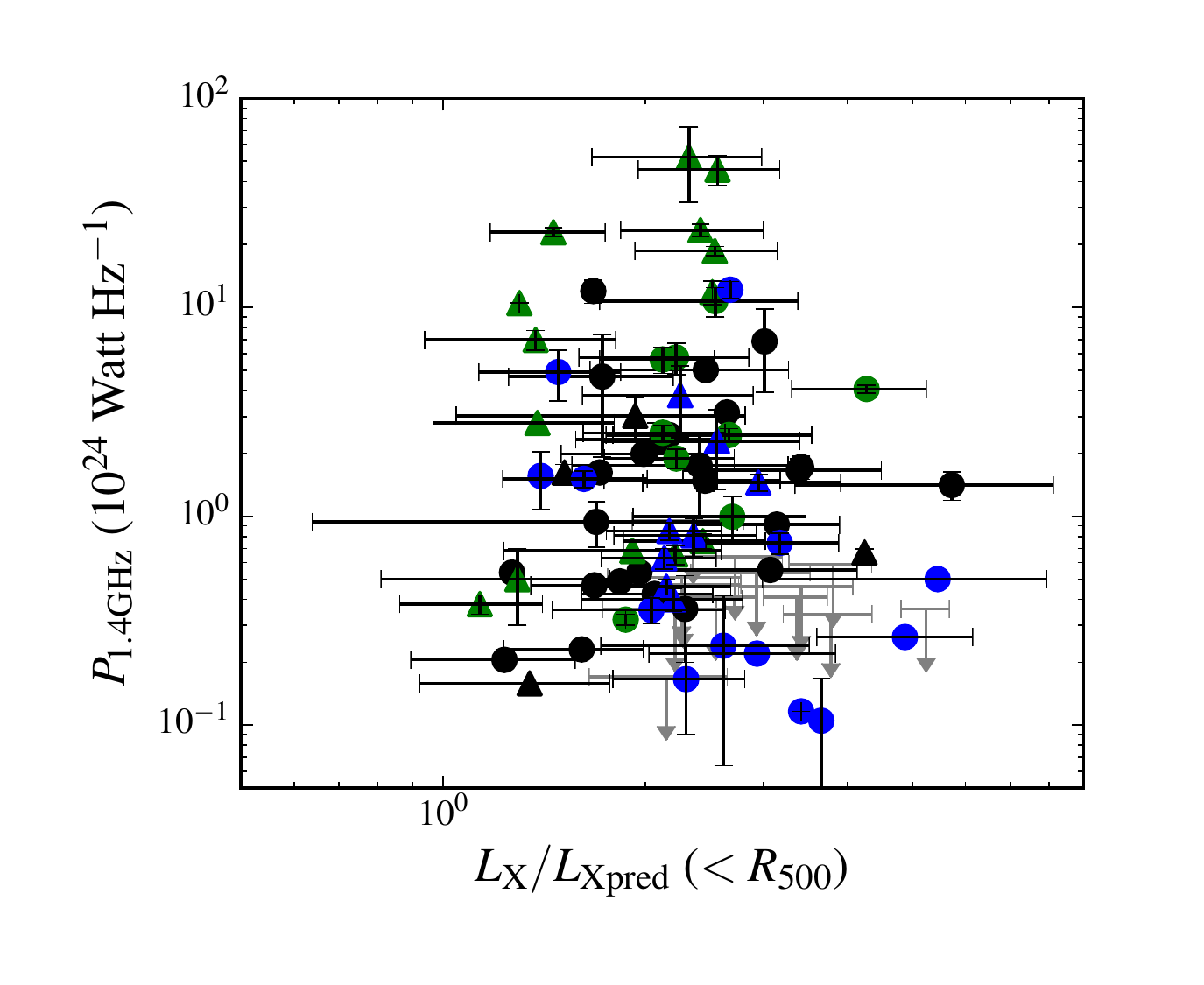} \\ \end{tabular}
\caption{The monochromatic 1.4 GHz radio power versus the ratio between the
measured X-ray luminosity and the
predicted X-ray luminosity from the SZ-derived cluster mass using the $LM$ scaling relations of \citet{mant10} (\emph{left} panel) and \citet{gile17} (\emph{right} panel). The green symbols denote the systems with flatter-spectrum ($\alpha<1.5$) haloes, the blue symbols denote the steeper-spectrum ($\alpha>1.5$) haloes, and the black symbols denote the systems that lack spectral information in the literature. The triangle symbols denote the RHs plus relic (radio shock) systems, and the circle symbols denote the systems that do not have relic emission.
In grey we overplot the upper limits from the merging systems without detected RH emission \citep{cass13,kale15}.}
\label{F:Phalo_vs_Lcalc_1}
\end{figure*}

To investigate the origins of the scatter seen in Figure \ref{F:Phalo_vs_L_M} further, we can search for relations
between the properties of the RH and the degree to which the X-ray luminosity
has been boosted (or suppressed). To this end, we calculate the ratio between
the measured X-ray luminosity, $L_{\rm X}(R<500)$, and the X-ray luminosity
predicted from the SZ derived mass, $L_{\rm Xpred}(R<500)$. To calculate the
latter, we use the well-known scaling between the cluster luminosity and cluster
mass. There is a large literature on the cluster luminosity-mass ($LM$) scaling
relation and its form
\citep[e.g.;][]{reip02,alle03,prat09,vikh09,mant10,gile17}, with the slope of
the relation varying across studies from $\sim 1.3$ \citep{alle03,mant10} to
$\sim 1.6$ and above \citep{vikh09,prat09,mant16,gile17}.
We use two recent determinations to
calculate $L_{\rm Xpred}$: the relation of
\citet{mant10}, which has a $LM$ normalization of 0.82 $\pm$
0.11 and a $LM$ slope of 1.29 $\pm$0.07 \citep[see Table 7 of][]{mant10}, and the
relation of \citet{gile17}, which has a slope of 1.92 $\pm$ 0.24 \citep[see
Table 4 and Table B1 of][]{gile17}. Both of these relations include corrections
for sample biases that account for the tendancy of X-ray selected samples to
preferentially include clusters that have higher luminosities than typical for a
given mass \citep[for a discussion, see][]{gile17}. We plot the halo radio power
versus the ratio between the measured X-ray luminosity and that calculated using
these scaling relations in Figure \ref{F:Phalo_vs_Lcalc_1}.

We find that the measured X-ray luminosity is higher on average by a factor of
$\sim 1.5$--2 than that predicted by the $LM$ relations, implying that clusters
with RHs tend to be overluminous for a given mass relative to the average over
all clusters. One explanation for this overluminosity is that clusters with RHs
are preferentially caught in a state soon after a major merger has occurred,
when the X-ray luminosity is expected to be boosted \citep[e.g.,][]{donn13}. An
alternative explanation is that our sample is biased towards overluminous
systems, for example due to selection effects. Our sample is largely based on
X-ray selected samples, so a sample bias of this kind is possible. Samples of RH
systems selected on other properties, such as the cluster mass, would be very
useful in understanding whether the overluminosity we observe is an intrinsic
property of RH systems or not \citep{cuci15,kale18}.

We note that the values of $L_{\rm Xpred}(R<500)$ calculated using the scaling relation of \citet{mant10}
are $\sim 1.5$ times higher than those calculated using that of \citet{gile17} for
our sample. This difference is mainly due to the differing slopes between the
two relations and the fact that our sample is comprised mostly of clusters with
masses below $\sim 10^{15}$~M$_{\odot}$, where this difference in slope has the
greatest effect. There is also an additional smaller systematic offset of $\approx
1.1$ between the luminosities used in \citet{mant10} and \citet{gile17} that we
do not correct for \citep[for details see][]{gile17}. One consequence of this
difference is that, for the \citet{mant10} scaling relation, the ratio $L_{\rm
X}/L_{\rm Xpred}$ falls below unity for a number of systems. Such ratios are not
expected in simulations until late in the merging process \citep[e.g.,][]{donn13}, well
after the RH should have faded away. Therefore, the low ratios could be interpreted as
indirect support for the higher slope of the $LM$ relation of \citet{gile17} (which
does not result in such low ratios). However, the low ratios could also occur as
a consequence of the intrinsic scatter about the $LM$ relation.

To investigate how the measured-to-predicted luminosity ratio relates to the
spectral properties of the radio halo, we separated the full sample into two
categories: systems with steeper spectral indices ($\alpha>1.5$)\footnote{The
values of $\alpha$ are listed in Table \ref{RH_table} and otherwise were taken from
\citet{cass13}, plus A2256 \citep[1.6,][]{bren08}, A2255 \citep[1.6,][]{pizz09},
RXC J1514.9-1523 \citep[1.6,][]{giac11c}, and A2034 \citep[1.7,][]{shim16}.}
and systems with flatter spectral indices
($\alpha<1.5$)\footnote{The values of $\alpha$ are listed in Table
\ref{RH_table} and otherwise are as follows: A2744 \citep[1.43,][]{pear17},
A2163 \citep[1.18,][]{fere04}, RXC J2003.5-2223 \citep[1.3,][]{giac09}, A520
\citep[1.12,][]{vacc14,hoag19a520}, A1758N \citep[1.2,][]{bott18}, A2219
\citep[0.9,][]{orru07}, A665 \citep[1.04,][]{fere04}, Coma
\citep[1.34,][]{kim90}, the Bullet cluster \citep[1.5,][]{shim14}, MACS
J0717.5+3745 \citep[1.4,][]{bona18}, El Gordo \citep[1.2,][]{lind14}, MACS
J1752.5+4440 \citep[1.33,][]{bona12}, and A3888 \citep[1.48,][]{shak16}.}. These
two subsamples are indicated by the different colors in Figure
\ref{F:Phalo_vs_Lcalc_1}.

We find that $P_{1.4{\rm ~GHz}}$ appears to be correlated with $L_{\rm X}/L_{\rm
Xpred}$ in the flatter-halo subsample when using the \citet{mant10} scaling
relation. To quantify the strength of this trend, we calculated the Spearman's
rank correlation coefficient. We find that the correlation is significant, with
a correlation coefficient of $0.70$ and a probability that the two quantities
are unrelated of $1 \times 10^{-4}$. However, there is no such correlation when
the relation of \citet{gile17} is used. If present, such a correlation would imply
that systems with powerful radio haloes are those for which the X-ray luminosity
is most affected (relative to the mass).

However, a difference between the two subsamples is evident in both panels
of Figure \ref{F:Phalo_vs_Lcalc_1}: flatter systems tend to have lower ratios of
measured-to-predicted luminosity and higher radio powers than steeper
systems (albeit with considerable overlap). It has been proposed that
a category of the steep-spectrum halos may be formed in low-turbulence mergers \citep{cass06}.
Since the radio power of the halo decreases as the merger evolves, at a given radio power
steep systems will tend to be observed at an earlier stage of the merger than
flatter ones. This expectation is consistent with the observed distribution of
steep halos in Figure \ref{F:Phalo_vs_Lcalc_1}, as systems observed at an
earlier stage are also expected to have a higher ratio  of measured-to-predicted X-ray luminosity
 than those observed at
later stages, when the X-ray luminosity has decreased. Therefore, the tendency
for steep-spectrum, low-power halo systems to have high ratios of measured-to-predicted X-ray luminosity is
broadly consistent with this scenario.

In support of this interpretation, the steep systems with the lowest RH power in
our sample are A3562 and A2811, which are also the lowest-mass systems in the
sample. As a result, mergers in these systems are expected to be less turbulent
than in high-mass systems \citep{cass06}. Other low-power, steep-spectrum RHs in
the plot are the recently identified RHs A1132, RXC J0142.0+2131, RXJ1720.1+2638 and
PSZ1G139.61+24.20 \citep{wilb18,savi19}, located in the lower-right corner.
These RHs were interpreted as likely having been created in lower-turbulence
merger events. They all have high ratios of measured-to-predicted X-ray luminosity, suggesting
they were caught in a stage that is fairly close to the core passage.

Therefore, the combination of the ratio of measured-to-predicted X-ray luminosity and the spectral
properties of the halo appears to be a general indicator of the merger stage.
Further support for this interpretation comes from the location of halo systems
with radio relics in the plot. We indicate such systems in Figure
\ref{F:Phalo_vs_Lcalc_1} (triangles): it is clear that systems with relics tend
to have low ratios of measured-to-predicted X-ray luminosity at a given halo radio power, especially relative to other
systems of the same spectral class (i.e., flat or steep). This tendency is in line with merger
simulations \citep[e.g.,][]{vazz12,ha18} that posit that relics are generally
found at a late stage of the merger ($\sim 1$ Gyr after core passage), when the
shock has propagated to large enough radii ($\sim 1$~Mpc) that the Mach number
is sufficiently high to efficiently create the relics. At these later stages, the X-ray
luminosity and halo radio power have decreased, and the relic systems therefore
tend to lie to the left of younger (non-relic) systems in Figure
\ref{F:Phalo_vs_Lcalc_1}.

Lastly, in Figure \ref{F:Phalo_vs_Lcalc_1} (\emph{right} panel), we plot
the RH upper limits from \citet{cass13} and \citet{kale15}. There are 20 such
systems in \citet{cass13} and two more systems in \citet{kale15} which have
masses derived from SZ observations. However, in 4 out of the 20 upper limits
systems from \citet{cass13} have been detected RH emission (e.g., A141, A2146,
A2261, and RXCJ0142.0+2131, see Table \ref{RH_table} for references), and as a
result, they do not belong to the upper limits class category. Furthermore, we
did not include in the RH upper-limit sample the strong cooling flow systems
without signs of merging activity (e.g., AS780, A3088, RXCJ1115.8+0129). As a
result, we have a sample for the  RH upper limits  of 13 systems  (see Table
\ref{Lx_table}). Figure \ref{F:Phalo_vs_Lcalc_1} shows
that systems with upper limits share the same region of the plot as the
steep-spectrum RHs, in line with steep-spectrum RH formation models
\citep[e.g.;][]{brun09,cass10a} that posit that some of these systems may have
faint, steep-spectrum haloes that remain undetected in current observations.

\subsection{The Candidate Radio Relic}\label{S:radio_relic}

One of the most prominent diffuse features in both the low- and high-resolution
LOFAR images of A959 is the linear feature, $\sim 400$~kpc in length and $\sim 125$~kpc
in width, located $\sim 800$~kpc to the south-east of the cluster core. The
location, orientation, and elongated, linear morphology of this feature strongly
resembles those of cluster radio relics. In support of this scenario, there are
no obvious optical counterparts that could explain the emission as being
associated with a radio galaxy.

As discussed in the introduction, radio relics are thought to be created in
merging systems, when electrons are accelerated or re-accelerated by the merger shocks. There are a number
of halo systems that show evidence of X-ray shocks associated with the
radio relic emission \citep[e.g. A521, Bullet, A754, El Gordo,
A2146;][]{giac06,giac08b,shim15,maca11,bott16a,russ10,hlav18}.
Such shocks, thought to be generated during the merger, are typically found to lie roughly
perpendicular to the
merger axis, often at the outskirts of the cluster. The complex distribution of
mass and galaxies in A959 makes it difficult to determine the merger axis,
but there is an elongation in the weak-lensing maps in the direction of the relic
that could indicate that the merger axis is along this line \citep{dahl03,bosc09}. In this case, the
putative relic meets many of the characteristics of known relics: it lies $\sim
1$~Mpc from the cluster center, and thus at the cluster outskirts; it is located
roughly along the merger axis; and its long axis is oriented perpendicular to
the merger axis. However, confirmation that it is a relic requires radio data at
higher frequencies to confirm that the spectral and polarization properties are that of a relic.

We measure the luminosity of the putative relic to be $P_{142{\rm ~MHz}} = (2.85
\pm 0.32) \times 10^{24}$~W~Hz$^{-1}$. We do not detect the relic in a
lower-resolution 322.7~MHz GMRT image (with a restoring beam of $39\arcsec
\times 50$\arcsec), implying a lower limit on the spectral index of $\alpha >
0.7$. There is no evidence in either the \emph{Chandra} or XMM-\emph{Newton}
images of a surface-brightness edge in the region of the relic that would be
indicative of a shock associated with it. However, both exposures have few
counts ($\lesssim 0.2$~counts~pixel$^{-1}$, where 1 pixel = 0.4919 arcsec on a
side for \emph{Chandra} and 1.1 arcsec on a side for XMM-\emph{Newton}) at this
location, and any edge produced by a typical shock would not be visible.
Therefore, deeper X-ray data are needed to confirm the presence of a shock at
this location.

\subsection{The Dark Clump}\label{S:dark_clump}

\citet{dahl03} identified a possible dark mass clump in their weak-lensing map
of A959. The clump, designated WL 1017.3+5931, lies to the south-west of
the cluster center and has little-to-no associated X-ray emission or
galaxy overdensity \citep{bosc09}. We can place limits on the X-ray
gas mass fraction in the clump using the XMM-\emph{Newton} data (we do not use
the \emph{Chandra} data for this purpose as they are shallower and therefore any
limits derived from them would be less constraining).

To this end, we measured the count rates in the exposure-corrected
XMM-\emph{Newton} images discussed in Section~\ref{S:xmm_analysis} in the
dark-clump and background regions shown in Figure~\ref{F:xray_images}. The
dark-clump region was chosen to encompass the majority of the mass peak found by
\citet{dahl03} while excluding the nearby X-ray point source and has a radius of
$r = 156$~kpc at the redshift of A959. For the background emission in the
region of the dark clump, which is comprised of the local background emission
from the main cluster and the instrumental background, we used the mean count
rate in an annulus centered on the cluster with inner and outer radii that match
those of the dark-clump region (see Figure~\ref{F:xray_images}).

In the dark-clump region, we measure an upper limit on the background-subtracted count rate, summed
over all three detectors, of $(1.6 \pm 5.4) \times 10^{-6}
$~count~s$^{-1}$~pixel$^{-1}$. Therefore, we do not detect significant excess
emission from the dark clump. To place limits on the density of X-ray gas in the
clump, we obtained predicted count rates from \textsc{pimms} (the Portable,
Interactive Multi-Mission Simulator\footnote{See
\url{https://heasarc.gsfc.nasa.gov/cgi-bin/Tools/w3pimms/w3pimms.pl}}), using
the APEC thermal plasma model with a temperature of 3 keV and an abundance of
0.3 times the solar abundance. We adjusted the normalization of the APEC model to match
the upper limit on the count rate in the dark-clump region,
accounting for the encircled-energy fraction of the point spread function for
this region ($\approx 0.8$).\footnote{See
\url{https://heasarc.gsfc.nasa.gov/docs/xmm/uhb/offaxisxraypsf.html}.} The upper limit is defined as three
times the uncertainty of the background count rate in the region (i.e., the
3-$\sigma$ upper limit, $1.6 \times 10^{-5} $~count~s$^{-1}$~pixel$^{-1}$).

From the resulting normalization, and assuming the gas fills a sphere with
uniform density, we find the limit on the electron density in the dark clump of
$n_{\rm{e}} < 3\times 10^{-4}$~cm$^{-2}$. This density implies a total gas mass
of $M_{\rm gas} < 1.2 \times 10^{11}$~$M_{\odot}$ (assuming $n = 2n_{\rm{e}}$).
\citet{dahl03} report a total mass for the dark clump of $M_{\rm tot} = 1.2
\times 10^{14}$~$M_{\odot}$ within a radius of $r = 230$~kpc (adjusted to our
adopted cosmological parameters). Again assuming spherical geometry and a
uniform density to adjust for the slightly different radii ($r = 230$~kpc for
the total mass and $r = 156$~kpc for the gas mass), we find the upper limit on
the gas mass fraction within $r = 156$~kpc to be $f_{\rm gas} < 1.7 \times
10^{-3}$. However, the total mass estimate should be treated with caution since
only weak lensing data were used \citep[see A2744,][]{jauz16}.
Nevertheless, the low gas mass fraction implies the clump, if real, was
efficiently stripped of its X-ray gas, similar to other, X-ray gas-poor mass
concentrations \citep[e.g.,][]{jee14,wang16,jee16}.

In addition to A959, there are a number of other systems in which a
dark clump has been reported, e.g., A2744 and A520 \citep[see also the review of][]{witt18}. However, for A2744 the dark
clump reported in \citet{mert11} was not confirmed by \citet{jauz16}, who used
both weak- and strong-lensing data. For A520 the results are also controversial, with
some works detecting a dark clump \citep{mahd07,okab08,jee12,jee14} and others
finding no significant detection \citep{clow12,peel17}.

\section{Summary \label{S:summary}}

Using LOFAR Two-meter Sky Survey (LoTSS) data we have identified a radio halo
and likely radio relic in A959. The RH has a flux at 144 MHz of $S_{143.7{\rm
~MHz}} = 0.094 \pm 0.014$~Jy. Using the measured flux at 1400 MHz for all
diffuse emission from \citet{owen99}, we found a spectral index for the RH of
1.48$^{+0.06}_{-0.23}$. Additionally, we report the detection of a likely radio
relic in A959, $\sim 400$~kpc in length and $\sim 125$~kpc in width, located
$\sim 800$~kpc to the south-east of the cluster core. There is no indication of
a surface brightness edge in the actual \emph{Chandra} and XMM-\emph{Newton}
data, but both have very few counts at the relic location ($\lesssim
0.2$~counts~pixel$^{-1}$). Deeper X-ray data will be required to search for
shocks in the ICM at the relic location.

We also examined the putative dark clump WL 1017.3+5931 for which no associated
galaxy concentration has been identified \citep{dahl03}. Using the XMM X-ray
data and the total mass from \citet{dahl02}, we placed limits on the X-ray gas
mass fraction in the clump. We find the upper limit on the gas mass fraction
within $r = 156$~kpc to be $f_{\rm gas} < 1.7 \times 10^{-3}$, implying
efficient stripping of the gas. However, this value (and the existence of the
clump itself) should be treated with caution, since only weak-lensing data were
used to measure the mass distribution, which consequently could have significant
uncertainties \citep[see, e.g., A2744 and A520,][]{jauz16,clow12,peel17}.

To place the diffuse radio emission in A959 in context, we collected all known
RH detections from the literature (80 systems in total) and added A959 to plots
between the non-thermal and thermal power \citep[e.g.,][]{cass13,mart16} of this
full RH sample. We find that the RH of A959 falls close to the scaling relations
of \citet{cass13}. As previously reported
\citep{brun09,basu12,cass13,kale15,cuci18}, there is a large scatter in these
scaling relations. This scatter may be partly explained as being due to
evolution in the radio and X-ray luminosities during the merger
\citep[e.g.,][]{rick01,ritc02,rand02,donn13}.

To investigate such evolution, we examined how the halo radio power relates to
the ratio between the measured X-ray luminosity and that predicted from the SZ
cluster mass using the cluster $LM$ scaling relations of \citet{mant10} and
\citet{gile17}, and we summarize the results below:

$   \bullet$ \noindent{\emph{ We find evidence that the flat-spectrum haloes occur in
systems with lower X-ray luminosity ratios and  higher halo radio powers, while the
steep-spectrum haloes tend to occur in systems with higher X-ray luminosity
ratios and lower radio powers.}}
We argue that this result is consistent with the
expectations of turbulent re-acceleration models of halo formation
\citep[e.g.,][]{brun09,cass10a}, where the halo spectral steepness is strongly
influenced by the level of turbulence generated by the merger. Specifically,
in these models, steep-spectrum haloes are expected to be created preferentially
in low-turbulence mergers \citep{cass06}, where the expected lifetime of the
halo is short. The short lifetimes imply that such systems \citep[e.g;
RXJ1720.1+2638,][]{savi19} are more likely to be observed at an earlier stage
of the merger than the systems with longer-lived, flatter haloes.

$   \bullet$ \noindent{\emph{We also find evidence that the RH systems with radio relics
have lower measured-to-predicted X-ray luminosities than similar non-relic
systems.}}
This finding is consistent with simulations of relics
\citep[e.g.,][]{vazz12,ha18}, which find that relics tend to be observed in the
cluster outskirts at the later stages of the merger, when the X-ray luminosity is
expected to have decreased significantly.

We therefore posit that the
combination of measured-to-predicted X-ray luminosity and the spectral
properties of the RH is a general indicator of the merger stage, in line with
simulations \citep{ritc02,donn13}.

\begin{table*}
\small
\begin{minipage}{145mm}
\caption{Radio Halo properties for the additional systems}
\label{RH_table}
\scalebox{0.95}{
\begin{tabular}{@{}lccccccc}
\hline
&    & $M_{\rm SZ 500}$ & Freq. & Flux density & $\alpha$$^c$ &
$P_{\rm 1.4GHz}$$^d$   \\
System$^a$$^,$$^b$  & $z$    &
 (10$^{14}$ M$_{\odot}$) & (MHz) & (mJy) & & (10$^{24}$ W Hz$^{-1}$)
  \\
\hline
A959                    & 0.288     & 5.08 $\pm$ 0.47 (2) & 143.7 & 94 $\pm$ 14 (6)  & 1.48 (6) & 0.68 (31)  \\
\hline
\hline
A141$^{U}$               & 0.23     & 4.48  $\pm$ 0.7 (5)  & 168   & 110 $\pm$ 11 (17)    & $>$ 2.1 (17) & $<$ 0.5  \\
A399                     & 0.0718   & 5.29 $\pm$ 0.34 (5)  & 1400  & 16 $\pm$ 2 (29)      & \ldots &  0.21 $\pm$ 0.03   \\
A401                     & 0.0737   & 6.84 $\pm$ 0.32 (4)  & 1400  & 17 $\pm$ 1 (8)      & \ldots & 0.23 $\pm$ 0.01    \\
A523                     & 0.104    & \ldots               & 1400  & 72 $\pm$ 3 (23)      & \ldots & 2.04 $\pm$ 0.08   \\
A800                     & 0.2223   & \ldots               & 1400  & 10.6 (24)            & \ldots &  1.64   \\
A851*                    & 0.4069   & \ldots               & 1400  & 3.7 $\pm$ 0.3 (21)   & \ldots & 2.41 $\pm$ 0.20     \\
A1132$^{U}$              & 0.1369   & 5.87 $\pm$ 0.22 (4)  & 145   & \ldots               & 1.75 (41) & 0.17 $\pm$ 0.08 (41)  \\
A1451                    & 0.199    & 7.16 $\pm$ 0.32 (3)  & 1500  & 5.0 $\pm$ 0.6 (14)   & $>$1.3 (14) & $<$0.66 $\pm$ 0.07 \\
A1550                    & 0.254    & 5.55 $\pm$ 0.54 (5)  & 1400  & 7.7 (24)             & \ldots &  1.62  \\
A1682$^{U}$              & 0.226    & 5.70 $\pm$ 0.35 (4)  & 240   & 46 $\pm$ 4 (39)      & 1.7 (28) & 0.40 $\pm$ 0.05    \\
A2061                    & 0.0777   & 3.32 $\pm$ 0.27 (5)  & 300   & 270 $\pm$ 2 (33)     & \ldots &  0.55 $\pm$ 0.01   \\
A2065                    & 0.073    & 4.30 $\pm$ 0.26 (5)  & 1400  & 32.9 $\pm$ 11 (18)   & \ldots & 0.48 $\pm$ 0.02   \\
A2069                    & 0.116    & 5.45 $\pm$ 0.37 (5)  & 1400  & 28.8 $\pm$ 7.2 (18)  & 0.93 (16) & 1.00 $\pm$ 0.02    \\
A2142                    & 0.089    & 8.77 $\pm$ 0.21 (4)  & 1400  & 23 $\pm$ 2 (40)      & \ldots & 0.47 $\pm$ 0.04  \\
A2218                    & 0.1756   & 6.59 $\pm$ 0.164 (4) & 1400  & 4.7 (20)             & \ldots & 0.43   \\
A2261$^{U}$              & 0.224    & 7.78 $\pm$ 0.30 (4)  & 1400  & 4.37 $\pm$ 0.35 (35) & 1.7 (34) & 0.75 $\pm$ 0.06    \\
A2811$^{U}$              & 0.1079   & 3.65  $\pm$ 0.24 (4)  & 168   & 80.7 $\pm$ 16.5 (17) & $>$1.5 (17) & $<$ 0.11  $\pm$   0.06  \\
A3562$^{U}$              & 0.049    & 2.3 (3)              & 1400  & 20 (37) & 1.56 (19,37)   & 0.12    \\
ACT-CLJ0256.5+0006       & 0.363    & 5.0 $\pm$ 1.2 (1)    & 610   & 5.6 $\pm$ 1.4 (27)    & \ldots & 0.94  $\pm$ 0.23   \\
AS1121*                  & 0.358    & 7.19 $\pm$ 0.45 (4)  & 168   & 154 $\pm$ 48 (17)    & \ldots & 4.66  $\pm$   2.75   \\
CIZA J0638.1+4747        & 0.174    &  6.65 $\pm$ 0.34 (3) & 1500  & 3.3 $\pm$ 0.2 (14)   & $>$1.3 (14) & $<$ 0.32 $\pm$ 0.02 \\
CIZA J2242.8+5301*       & 0.192    & \ldots               & 145   & 346 $\pm$ 64 (25)     & 1.03 (25) & 3.1 $\pm$  1.0    \\
CL0217+70                & 0.0655   & \ldots               & 1400  & 58.6 $\pm$ 0.9 (12)  & \ldots & 0.61 $\pm$ 0.09    \\
CL1446+26*               & 0.370    & \ldots               & 1400  & 7.7 (24)             &     \ldots      & 3.57 \\
H1821+643                & 0.332    & 6.78 $\pm$ 0.27 (4)  & 1665  & 19.9 $\pm$ 0.5 (11)  & 1.1 (11) &  4.07 $\pm$ 0.17  \\
MACS J0416.1-2403$^{U}$  & 0.393    & \ldots               & 1500  & 1.58 $\pm$ 0.13 (30) & 1.6 (30) & 1.16 $\pm$   0.09   \\
MACS J0417.5-1154$^{U}$  & 0.443    & 12.25 $\pm$ 0.55 (4) & 1575  & 10.6 $\pm$ 1.0 (32)  & 1.72 (32) & 12.15 $\pm$ 1.15    \\
MACS J2243.3-0935        & 0.44     & 9.99 $\pm$ 0.44 (1)  & 610   & 10.0 $\pm$ 2.0 (13,32)   & \ldots & 2.41 $\pm$ 0.28   \\
PLCKG004.5-19.5          & 0.516    & 9.42 $\pm$ 0.94  (5) & 610   & 1.2  $\pm$ 0.5 (7) & 1.2 $\pm$ 0.4 (7)  & 0.5 $\pm$ 0.2 \\
PLCKG287.0+32.9          & 0.39     & 14 (2)               & 150   & 314 (10) & 1.28 (10) & 10.5  \\
PSZ1G018.75+23.57        & 0.089    & 3.97 $\pm$ 0.30 (4)  & 1860  & 48.3 $\pm$ 2.5 (9)  & \ldots &  1.42 $\pm$ 0.07     \\
PSZ1G108.18-11.53        & 0.335    & 7.74  $\pm$ 0.60 (4) & 1380  & 6.8 $\pm$ 0.2 (15)   &  1.4 $\pm$ 0.1 & 2.8 $\pm$ 0.1  \\
PSZ1G139.61+24.20$^{U}$  & 0.27     & 7.09 $\pm$ 0.60 (5)  & 144   & \ldots               &  $>$ 1.7 (34) & $<$ 0.22 (34) \\
RXC J0142.0+2131$^{U}$   & 0.28     & 5.98 $\pm$ 0.60 (4)  & 144   & 32 $\pm$  6 (34)     &  $>$ 1.6 (34)   &  $<$ 0.24 \\
RX J0603.3+4214*        & 0.225    & 10.72 $\pm$ 0.49 (4) & 1500  & 46 $\pm$ 5 (36)      & 1.08 (36) & 7.00 $\pm$ 0.76   \\
RXC J1314.4-2515         & 0.228    & 6.15 $\pm$ 0.7 (3)   & 610   & 10.3 $\pm$ 0.3 (38)  & \ldots & 0.67  $\pm$ 0.03    \\
RXJ1720.1+2638$^{U}$     & 0.164    & 5.90 $\pm$  0.34 (4)  & 144  &  \ldots              &  $>$ 1.5 (34) & $<$ 0.264 (34) \\
Triangulum Aus.          & 0.051    & 7.94 $\pm$ 0.15 (4)  & 1330  & 92 $\pm$ 5 (9)      & \ldots & 0.54 $\pm$ 0.03     \\
ZwCL2341.1+0000          & 0.27     & 5.18 $\pm$ 0.44 (4)  & 1400  & 10 (22)              &  \ldots  & 0.16   \\
ZwCL5247*$^{U}$      & 0.229    & 5.88  $\pm$ 0.40 (4) & 1400  & 2.0 $\pm$ 0.3 (26)   & 1.7 (26) & 0.35 $\pm$ 0.05   \\
\hline
\end{tabular}
}

References:
SZ References: (1) \citet{hass13}; (2) \citet{plan13}; (3) \citet{plan15}; (4) \citet{plank16}; (5) SZ-Cluster Database (see \url{http://szcluster-db.ias.u-psud.fr}).
Radio References:
(6)   this work;
(7)  \citet{albe17};
(8) \citet{bacc03}
(9) \citet{bern16};
(10) \citet{bona14};
(11)  \citet{bona14a};
(12)  \citet{brow11};
(13) \citet{cant16};
(14)  \citet{cuci18};
(15)  \citet{deGa15};
(16) Drabent et al. in press;
(17) \citet{duch17};
(18) \citet{farn13};
(19) \citet{giac05};
(20)  \citet{giov00};
(21) \citet{giov09};
(22)   \citet{giov10};
(23) \citet{gira16};
(24) \citet{govo12};
(25) \citet{hoag17};
(26) \citet{kale15};
(27) \citet{know16};
(28) \citet{maca13};
(29)  \citet{murg10a};
(30) \citet{ogre15};
(31) \citet{owen99};
(32) \citet{pare17};
(33) \citet{rudn09};
(34) \citet{savi19};
(35) \citet{somm17};
(36)  \citet{vanW16b};
(37)  \citet{vent03};
(38)  \citet{vent07};
(39) \citet{vent13};
(40) \citet{vent17};
(41) \citet{wilb18}.

$^a$The radio halo systems (taken from the literature) that are not present in the
\citet{cass13} and \citet{mart16} samples.  The asterisk marks
systems with alternative names: A851 (CL0939+47); AS1121 (SPT-CL J2325-4111); CL1446+26 (ZwCL1447+2619); CIZA J2242.8+5301
(the ''Sausage'' cluster), ZwCL5247 (RXC J1234.2+0947), RXC J0603.3+4214 (the ''Toothbrush'' cluster). The 'U' marks the systems with steep-spectrum RHs ($\alpha>1.5$).  \\
$^b$However, there are some candidate haloes that are not present here, e.g.; A2680, A2693, AS84, RXC J2351.0-1954, GMBCG J357.91841-08.97978 \citep{duch17}; A2552, ZwCL1953 \citep{kale15}. \\
$^c$Spectral index from the literature.
\\
$^d$Radio luminosity at 1.4 GHz using the spectral indices from the literature when available and adopting $\alpha$=1.3 otherwise.  \\
\end{minipage} \end{table*}

% second table

\begin{table*}
\small
\begin{minipage}{145mm}
\caption{X-ray Luminosity values for the total sample }
\label{Lx_table}
\scalebox{0.95}{
\begin{tabular}{@{}lcccccccc}
\hline
&  &  $L_{\rm X 500}[0.1-2.4 keV]$$^b$   &  $L_{\rm Xpred}^{\rm M}$$^c$ & $L_{\rm Xpred}^{\rm G}$$^d$   &   \\
System$^a$ & $z$  & (10$^{44}$ erg s$^{-1}$)  & (10$^{44}$ erg s$^{-1}$) & (10$^{44}$ erg s$^{-1}$)  & Relics$^e$
  \\
\hline
A959                    & 0.288  & 3.24 $\pm$ 0.46 (1) & 3.80 $\pm$ 1.02 & 1.69 $\pm$  0.55 &  (1) \\
\hline
\hline
A141$^{U}$              & 0.23    & 6.82 $\pm$ 0.27  (5) & 3.01  $\pm$ 1.06 & 1.25 $\pm$ 0.56 &  \ldots  \\
A399                    & 0.0718  & 1.82 $\pm$ 0.042 (18)  & 3.12 $\pm$ 0.73 & 1.47 $\pm$ 0.40 &  \ldots  \\
A401                    & 0.0737  & 3.85 $\pm$ 0.05 (11)  & 4.33 $\pm$ 0.90 & 2.39 $\pm$ 0.56 &  \ldots  \\
A523                    & 0.104   & 0.91 (12) & \ldots & \ldots & \ldots   \\
A800                    & 0.2223  & 2.72 (12) &  \ldots & \ldots & \ldots  \\
A851*                   & 0.4069  & 4.91 (12) & \ldots   &  \ldots & \ldots    \\
A1132$^{U}$             & 0.1369  & 4.4 $\pm$ 0.1 (16)   & 3.83 $\pm$ 0.76 & 1.91  $\pm$ 0.42 &  \ldots \\
A1451                   & 0.199   & 6.61 (12)  & 5.3 $\pm$ 1.10 & 2.98 $\pm$ 0.70 &  (26) \\
A1550                   & 0.254   & 3.32 (12)  & 4.09 $\pm$ 1.13 & 1.94  $\pm$ 0.65 &  \ldots \\
A1682$^{U}$             & 0.226   & 4.36 $\pm$ 0.11  (6)  & 4.09 $\pm$ 0.94 & 1.98 $\pm$  0.53 &   (51)  \\
A2061                   & 0.0777  & 1.86 $\pm$ 0.30 (18) & 1.72 $\pm$ 0.44  & 0.61 $\pm$ 0.18 &    \ldots  \\
A2065                   & 0.073   & 1.82 (14)  &  2.40 $\pm$ 0.55 & 0.99 $\pm$ 0.27 &  \ldots \\
A2069                   & 0.116   & 4.27 $\pm$ 0.69 (18) &  3.34 $\pm$ 0.70 & 1.58  $\pm$ 0.38 &  \ldots  \\
A2142                   & 0.089   & 6.65 $\pm$ 0.05 (11) & 6.11 $\pm$ 1.10 & 3.95 $\pm$ 0.77 &  \ldots \\
A2218                   & 0.1756  & 5.1 $\pm$ 0.5 (9)    & 4.62  $\pm$ 0.84 & 2.47 $\pm$ 0.49 &  \ldots  \\
A2261$^{U}$             & 0.224   & 11.38 $\pm$ 0.13 (6)  &  6.09 $\pm$ 1.21 & 3.58 $\pm$ 0.80 &  \ldots \\
A2811$^{U}$                   & 0.1079  & 2.73 (13)   & 2.01 $\pm$ 0.47  & 0.75 $\pm$  0.21 &  \ldots \\
A3562$^{U}$                   & 0.049   & 0.997 $\pm$ 0.032  (11) & 1.04 $\pm$ 0.16 & 0.29 $\pm$ 0.04 &   \ldots \\
ACT-CLJ0256.5+0006      & 0.363   & 3.01 $\pm$ 0.36 (8)  & 4.08 $\pm$ 1.87 & 1.78 $\pm$ 1.09 &  \ldots \\
AS1121*                 & 0.358   & 6.14 $\pm$ 0.37 (3)  & 6.48 $\pm$ 1.49 & 3.56 $\pm$ 0.96 &  \ldots \\
CIZA J0638.1+4747       & 0.174   & 4.72 (12)      &  4.69 $\pm$ 1.01 & 2.52 $\pm$ 0.63 &  \ldots \\
CIZA J2242.8+5301*      & 0.192   & 7.7 $\pm$ 0.1  (7)  &  \ldots & \ldots &  (7,43)   \\
CL0217+70               & 0.0655  & 0.575 $\pm$ 0.202 (18) &  \ldots & \ldots & \ldots    \\
CL1446+26*              & 0.370   & 3.42 (16)   &  \ldots & \ldots & \ldots  \\
H1821+643               & 0.332   & 13.18 $\pm$ 0.03 (18) &  5.81 $\pm$   1.17 & 3.08 $\pm$ 0.70 &  \ldots \\
MACS J0416.1-2403$^{U}$ & 0.393   & 9.14 $\pm$ 0.10 (10) &  \ldots & \ldots & \ldots    \\
MACS J0417.5-1154$^{U}$ & 0.443   & 29.1 (12)  &  14.34 $\pm$ 2.98 &  10.86 $\pm$ 2.57 &  \ldots  \\
MACS J2243.3-0935       & 0.44    & 15.2 $\pm$ 0.8 (9) &  10.98 $\pm$ 2.27 & 7.32 $\pm$ 1.72 &  (24)  \\
PLCKG004.5-19.5         & 0.516   & 9.2 $\pm$ 1.4 (2) &  11.23 $\pm$ 3.13 & 7.13 $\pm$ 2.43  &  (2)  \\
PLCKG287.0+32.9         & 0.39    & 17.2 $\pm$  0.11 (13) &   15.92 $\pm$ 2.39 & 13.23 $\pm$ 1.98 &  (20,21) \\
PSZ1G018.75+23.57       & 0.089   & \ldots    &  2.20 $\pm$ 0.54 & 0.86 $\pm$ 0.26 &  \ldots  \\
PSZ1G108.18-11.53       & 0.335   &  5.52 $\pm$ 0.23 (1)*   &   6.92 $\pm$ 1.73 &  3.99 $\pm$ 1.19 &  (27)  \\
PSZ1G139.61+24.20$^{U}$ & 0.27    &   9.22 $\pm$ 0.23 (1)*  &   5.71 $\pm$ 1.46  & 3.15 $\pm$ 0.97  &  \ldots \\
RXC J0142.0+2131$^{U}$  & 0.28    & 6.0 $\pm$ 0.1 (5)   & 4.63 $\pm$ 1.30 & 2.29 $\pm$ 0.79 &  \ldots  \\
RX J0603.3+4214*       & 0.225   & 9.12 $\pm$ 1.90 (17) & 9.21 $\pm$  1.93 & 6.64 $\pm$ 1.58 &    (45,48)  \\
RXC J1314.4-2515        & 0.228   & 9.89 (12)            &  4.60 $\pm$  1.37 & 2.33  $\pm$ 0.86 &  (50) \\
RXJ1720.1+2638$^{U}$   & 0.164    & 9.69 $\pm$ 0.10 (6)  & 3.98 $\pm$ 0.89   & 1.99 $\pm$ 0.52  &   \ldots   \\
Triangulum Aus.         & 0.051   & 3.97 $\pm$ 0.08  (11)   &   5.14 $\pm$ 0.77 & 3.13  $\pm$ 0.47 &  \ldots  \\
ZwCL2341.1+0000         & 0.27    &   2.32 $\pm$ 0.06 (1)*    &  3.81 $\pm$ 0.99 &  1.72 $\pm$ 0.54 &   (42)  \\
ZwCL5247*$^{U}$         & 0.229   & 4.3 $\pm$ 0.3 (9)    & 4.26 $\pm$ 1.01 & 2.10 $\pm$  0.59 &  \ldots \\
\hline
%\multicolumn{4}{c}{The systems from \citet{cass13} sample}\\
%\hline
A2744                   & 0.307 & 14.73 $\pm$ 0.24 (5)  & 8.76 $\pm$ 1.84  & 5.80 $\pm$ 1.38 &  (31,36)   \\
A209                    & 0.206 & 7.62 $\pm$ 0.48 (5)  & 6.31 $\pm$ 1.32    & 3.83 $\pm$ 0.91 &  \ldots  \\
A2163                   & 0.203 & 21.95 $\pm$ 0.33 (5) & 15.78 $\pm$ 2.84 & 15.03 $\pm$ 2.92 &  (28)\\
RXCJ2003.5-2323         & 0.317 & 9.17 $\pm$ 0.09 (5) & 6.40 $\pm$ 1.72 & 3.60 $\pm$ 1.18 &  \ldots  \\
A520                    & 0.203 & 7.81 $\pm$ 0.21 (5) & 5.26 $\pm$ 1.28  & 2.93 $\pm$ 0.96 &  \ldots \\
A773                    & 0.217 & 7.30 $\pm$ 0.57 (5)  & 5.35 $\pm$ 1.28 & 2.97 $\pm$  0.84 &   \ldots  \\
A1758N$^f$                  & 0.280 & 8.80 $\pm$ 0.16 (5) & 6.68 $\pm$ 1.60  & 3.96 $\pm$ 1.12 &  \ldots  \\
A2219                   & 0.228 & 14.78 $\pm$ 0.19 (5)  & 9.52 $\pm$ 1.71  & 6.97 $\pm$ 1.35 &  \ldots  \\
A521$^{U}$              & 0.248 & 8.28 $\pm$ 0.07 (5)  & 5.22 $\pm$  1.41 & 2.81 $\pm$ 0.92  &   (29,30) \\
A697$^{U}$              & 0.282 & 13.04 $\pm$ 0.61 (5) & 10.78 $\pm$ 2.26  & 8.04 $\pm$ 1.92 &  \ldots  \\
A1300$^{U}$             & 0.308 & 11.47 $\pm$ 0.37 (5) & 8.02 $\pm$ 1.92 & 5.08 $\pm$ 1.44 &  (51)  \\
CL0016+16               & 0.541 & 15.54 $\pm$ 0.28 (4) & 10.49 $\pm$ 2.82&  6.31 $\pm$ 2.06 &  \ldots   \\
A665                    & 0.182 & 8.30 $\pm$ 0.07 (5)  & 6.32 $\pm$  1.32  & 3.91 $\pm$ 0.93 &   \ldots   \\
A545                    & 0.154 & 6.31 $\pm$ 0.09 (5) & 2.67 $\pm$ 0.88  & 1.10 $\pm$ 0.46 &  \ldots   \\
Coma                    & 0.023 & 3.39 $\pm$ 0.03 (11) & 2.94 $\pm$ 0.62 & 1.39 $\pm$ 0.33 &  (19,41) \\
A2256$^{U}$             & 0.058 & 4.44 $\pm$ 0.02 (5) & 3.87 $\pm$ 0.70 &  2.04 $\pm$ 0.40 &  (23,25,35,46) \\
Bullet*                  & 0.296 & 22.54 $\pm$ 0.52 (5) & 11.98 $\pm$  2.51 &  9.32 $\pm$ 2.22 & (38,39)   \\
A2255$^{U}$             & 0.081 & 3.31 $\pm$ 0.03 (5) & 3.03 $\pm$ 0.64 & 1.40  $\pm$ 0.33 &  (32,37)  \\
A2319                   & 0.056 & 7.87 $\pm$ 0.08 (5) & 5.68 $\pm$ 1.02 & 3.63 $\pm$ 0.70 &  \ldots  \\
MACSJ0717.5+3745        & 0.548 & 24.05 $\pm$ 0.22 (5) & 14.68 $\pm$ 3.51 & 10.34 $\pm$ 2.92 &  (22,49)  \\
A1995                   & 0.319 & 6.03 $\pm$ 0.08 (5) & 3.99 $\pm$  1.07  & 1.78  $\pm$  0.58 &  \ldots   \\
MACSJ1149.5+2223$^{U}$  & 0.544 & 15.50 $\pm$ 0.29 (5) & 10.23 $\pm$  2.75 & 6.06  $\pm$  1.98 &  (4) \\
PLCKG171.9-40.7$^{U}$   & 0.270 & 11.28 $\pm$ 0.02 (13) & 10.31 $\pm$ 2.16  & 7.60 $\pm$  1.81 &  \ldots   \\
A754$^{U}$              & 0.054 & 4.75 $\pm$ 0.033 (5) &   4.09 $\pm$ 0.73 &  2.22 $\pm$ 0.43 &  (34)  \\
\hline
\end{tabular}
}
\end{minipage} \end{table*}

\setcounter{table}{1}
\begin{table*}
\begin{minipage}{145mm}
\caption{X-ray Luminosity values for the total sample -- cont.}
\label{Lx_table}
\scalebox{0.95}{
\begin{tabular}{@{}lcccccccc}
\hline
&  &  $L_{\rm X 500}[0.1-2.4 keV]$$^b$   &  $L_{\rm Xpred}^{\rm M}$$^c$ & $L_{\rm Xpred}^{\rm G}$$^d$  &  \\
System$^a$ & $z$  & (10$^{44}$ erg s$^{-1}$)  & (10$^{44}$ erg s$^{-1}$) & (10$^{44}$ erg s$^{-1}$)  & Relics$^e$
  \\
\hline
\multicolumn{4}{c}{The extra systems from \citet{mart16} sample}\\
\hline
A746                     & 0.2323 & 3.39 $\pm$ 1.19 (18)  & 3.78 $\pm$ 0.93 & 1.75 $\pm$ 0.52 &  (44) \\
A1351                    & 0.322  & 5.24 (12) &  5.84 $\pm$ 1.29  & 3.13 $\pm$ 0.80 &   \ldots \\
A1689                    & 0.1832 & 13.6 $\pm$ 1.2 (9)  &  6.77 $\pm$ 1.35 & 4.33 $\pm$ 0.97 &  \ldots  \\
A2034$^{U}$              & 0.113  & 4.0 $\pm$ 0.4 (9)   & 3.72 $\pm$ 0.75 & 1.86 $\pm$ 0.42 &  (40) \\
A2254                    & 0.178  & 4.79 (12)  & 3.76 $\pm$ 0.88  & 1.81 $\pm$ 0.50 &  \ldots  \\
A2294                    & 0.178  & 4.05 (12) & 4.11 $\pm$ 0.94 & 2.05 $\pm$ 0.55 &   \ldots \\
A3411                    & 0.1687 & 2.8 $\pm$ 0.1 (15) &  4.60 $\pm$ 0.96  & 2.47 $\pm$ 0.59 &  (47)  \\
A3888                    & 0.151  & 6.38 $\pm$ 0.25 (12)  & 5.06 $\pm$  0.10 & 2.86 $\pm$  0.63 &   \ldots \\
CIZAJ1938.3+5409         & 0.26   & 7.96 (12) &  6.06 $\pm$ 1.33  & 3.47 $\pm$   0.88 &  \ldots \\
El Gordo*                 & 0.87   & 35.48 $\pm$ 1.63 (18)  & 21.30 $\pm$ 4.40 & 13.84 $\pm$ 3.24 &  (33) \\
MACSJ0553.4-3342         & 0.431  & 17 (4) & 9.18 $\pm$ 1.98  & 5.64 $\pm$ 1.40 &  \ldots \\
MACSJ1752.0+4440         & 0.366  & 8.0 (4) & 6.03 $\pm$ 1.42 & 3.18  $\pm$ 0.89 &  (4) \\
PLCKG285.0-23.7          & 0.39   & 16.91 $\pm$ 0.27 (13)  & 8.22 $\pm$ 1.23  & 4.95 $\pm$ 0.74 &  \ldots \\
RXCJ0107.7+5408          & 0.1066 & 2.80  (12)  & 3.69 $\pm$ 0.80 & 1.85 $\pm$ 0.46 &  (44)  \\
RXCJ0949.8+1708          & 0.38   & 11.3 $\pm$ 2.3 (9)   & 7.89 $\pm$  1.93 & 4.69 $\pm$ 1.36 &   \ldots  \\
RXCJ1514.9-1523$^{U}$    & 0.226  & 6.43 (12) & 7.19 $\pm$ 1.56 & 4.60 $\pm$ 1.15 &  \ldots \\
\hline
\hline
\multicolumn{4}{c}{The upper limits systems from \citet{cass13} and \citet{kale15}}\\
\hline
A267                     & 0.230 & 5.94 $\pm$ 0.44 (5) & 3.51 $\pm$ 1.04  & 1.57 $\pm$ 0.58 &  \ldots \\
A781                     & 0.298 & 5.44 $\pm$ 0.14 (5) & 4.90 $\pm$ 1.24 & 2.45 $\pm$   0.74 &   \ldots  \\
A1423                    & 0.213 & 4.76 $\pm$ 0.38 (5) & 4.38 $\pm$ 1.12 &  2.21 $\pm$ 0.68 &  \ldots \\
A1576                    & 0.30  & 6.38 $\pm$ 0.14 (5) & 4.75 $\pm$ 1.20 & 2.34 $\pm$ 0.71 &   \ldots \\
A1722                    & 0.327 & 6.15 (12)           & 3.02 $\pm$ 0.87   & 1.17 $\pm$  0.41 &  \ldots \\
A2485                    & 0.247 & 3.07 $\pm$ 0.07 (5) & 3.19  $\pm$ 0.92 & 1.34 $\pm$ 0.48 &   \ldots \\
A2537                    & 0.297 & 4.54 $\pm$ 0.07 (5) & 4.27 $\pm$ 1.15 & 2.00  $\pm$ 0.65 &  \ldots \\
A2631                    & 0.278 & 8.62 $\pm$ 0.70 (5) & 6.01  $\pm$ 1.32 & 3.38 $\pm$ 0.85 &  \ldots \\
A2645                    & 0.251 & 4.13 $\pm$ 0.4 (5)  & 2.76 $\pm$  0.88 & 1.08 $\pm$ 0.44 &  \ldots \\
A2697                    & 0.232 & 7.29 $\pm$ 0.41 (5) & 4.35 $\pm$ 1.00  & 2.17 $\pm$  0.58 &  \ldots \\
RXCJ0439.0+0715          & 0.244 & 7.69 $\pm$ 0.58 (5) & 4.49 $\pm$ 1.31  & 2.25 $\pm$ 0.81 &   \ldots \\
RXJ2228.6+2037           & 0.418 & 11.71 $\pm$ 0.20 (5) & 8.36 $\pm$  1.84 & 4.96 $\pm$ 1.26 &  \ldots  \\
ZwCL7215                 & 0.2917 & 5.00 $\pm$ 0.19 (6)  & 3.82 $\pm$  1.15 & 1.71 $\pm$ 0.64 &  \ldots \\
\hline

\end{tabular}
}

References: X-ray References:
(1 ) this work;
(2)  \citet{albe17};
(3) \citet{birz17};
(4) \citet{bona12};
(5) \citet{cass13};
(6) \citet{gile17};
(7) \citet{hoag17};
(8) \citet{know16};
(9)  \citet{mant10};
(10) \citet{ogre15};
(11) \citet{ohar06};
(12) \citet{piff11};
(13) \citet{plan11};
(14) \citet{vikh09}
(15) \citet{vanW13}
(16) \citet{wilb18};
(17) \citet{wu99};
(18) \citet{yuan15}.
Relics references:
(19) \citet{ande84}
(20) \citet{bagc11}
(21) \citet{bona14}
(22) \citet{bona18}
(23) \citet{bren08}
(24) \citet{cant16}
(25) \citet{clar06a}
(26) \citet{cuci18}
(27) \citet{deGa15}
(28) \citet{fere01}
(29) \citet{giac06}
(30) \citet{giac08b}
(31) \citet{govo01}
(32) \citet{govo05}
(33) \citet{lind14}
(34) \citet{maca11}
(35) \citet{owen14}
(36) \citet{pear17}
(37) \citet{pizz09}
(38) \citet{shim14}
(39) \citet{shim15}
(40) \citet{shim16}
(41) \citet{thie03}
(42) \citet{vanW09a}
(43) \citet{vanW10}
(44) \citet{vanW11}
(45) \citet{vanW12a}
(46) \citet{vanW12}
(47) \citet{vanW13}
(48) \citet{vanW16b}
(49) \citet{vanW17}
(50) \citet{vent07}
(51) \citet{vent13}

$^a$Radio halo systems taken from the literature including those in the
\citet{cass13} and \citet{mart16} samples, and the systems with upper limits. The asterisk marks
systems with alternative names: Bullet (1E 0657-56), El Gordo (ACT-CL J0102-4915), with the others listed in Table \ref{RH_table}.
And, as in Table \ref{RH_table}. The 'U' marks the systems with steep-spectrum RHs ($\alpha>1.5$).\\
$^b$X-ray luminosity between 0.1-2.4 keV within R$_{500}$, except for CL1446+26, where only the bolometric X-ray luminosity was available in the literature \citep{wu99}; and for the systems from \citet{ohar06} and \citet{vikh09}, where the 0.5-2.0 keV energy band was used. For the systems marked with asterisk, since there were no available X-ray luminosities in the literature, we reduced the \emph{Chandra} X-ray data (ObsIDs 15139, 17490, 17213) ourselves, following the same reduction scheme described in Section \ref{S:chandra_analysis}. \\
$^c$The predicted X-ray luminosity between 0.1-2.4 keV within R$_{500}$ using the L-M scaling relations of \citet{mant10}. \\
$^d$The predicted X-ray luminosity between 0.1-2.4 keV within R$_{500}$ using the L-M scaling relations of \citet{gile17}.\\
$^e$The presence of relics (radio shocks) from literature.\\
$^f$ There is also a RH in A1758S \citep{bott18}. \\

\end{minipage} \end{table*}

\clearpage

\section*{Acknowledgements}

The scientific results reported in this article are based on data obtained with
 the International LOFAR Telescope (ILT), and archive data from \emph{Chandra} Data Archive and XMM-\emph{Newton} archive.
LOFAR \citep{vanH13} is the Low Frequency Array designed and constructed by ASTRON. It has observing, data processing, and data storage facilities in several countries, that are owned by various parties (each with their own funding sources), and that are collectively operated by the ILT foundation under a joint scientific policy. The ILT resources have benefitted from the following recent major funding sources: CNRS-INSU, Observatoire de Paris and Université d'Orléans, France; BMBF, MIWF-NRW, MPG, Germany; Science Foundation Ireland (SFI), Department of Business, Enterprise and Innovation (DBEI), Ireland; NWO, The Netherlands; The Science and Technology Facilities Council, UK.

The LOFAR reduction was done using PREFACTOR and FACTOR packages, and the X-ray data reduction has made using CIAO package provided by
Chandra X-ray Center (CXC), and SAS package for XMM-\emph{Newton} data.
The LOFAR group in Leiden is supported by the ERC Advanced Investigator program New-Clusters 321271. The authors thank the referee for the constructive comments, which improve the paper significantly.

\bibliographystyle{mn2e}
\bibliography{/Users/Laura/Documents/Bibliography/master_references}
\end{document}